\documentclass{article}
\usepackage{amsmath}
\usepackage{amsfonts}
\usepackage{amssymb}
\usepackage{amsthm}
\usepackage{ifpdf}
\usepackage{subfig}
\usepackage{wrapfig}
\usepackage{fullpage}
\usepackage{cite}
\usepackage{url}

\ifpdf

  \usepackage[pdftex]{epsfig}
  \usepackage[pdftex]{hyperref}

\else

    \usepackage[dvips]{epsfig}
    \newcommand{\href}[2]{#2}

\fi

\theoremstyle{definition}
\newtheorem{theorem}{Theorem}[section]
\newtheorem{lemma}[theorem]{Lemma}

\newtheorem{fact}{Fact}

\newtheorem{remark}[theorem]{Remark}
\newtheorem{question}[theorem]{Question}

\newcommand{\Z}{\mathbb{Z}}

\newcommand{\pfunc}[3]{#1 : #2 \dashrightarrow #3 }

\newcommand{\bval}[1]{[\![ #1 ]\!]}

\newcommand{\dom}{{\rm dom} \;}

\newcommand{\termasm}[1]{\mathcal{A}_{\Box}\left[#1\right]}

\newcommand{\fgg}[1]{G^\#_{#1}}

\begin{document}

\title{Reducing Tile Complexity for the Self-Assembly of Scaled Shapes Through Temperature Programming\footnote{This research was supported in part by National Science Foundation
Grants 0652569 and 0728806, and by NSF-IGERT Training Project in Computational Molecular Biology
Grant number DGE-0504304}}%

\author{Scott M. Summers \\ \\ Iowa State University \\ Department
of Computer Science \\ Ames, IA 50011, USA \\
\url{summers@cs.iastate.edu}}

\date{}

\maketitle

\begin{abstract}
This paper concerns the self-assembly of scaled-up versions of arbitrary finite shapes. We work in the multiple temperature model that was introduced by Aggarwal, Cheng, Goldwasser, Kao, and Schweller (\emph{Complexities for Generalized Models of Self-Assembly}, SODA 2004). The multiple temperature model is a natural generalization of Winfree's abstract tile assembly model, where the temperature of a tile system is allowed to be shifted up and down as self-assembly proceeds. We first exhibit two constant-size tile sets in which scaled-up versions of arbitrary shapes self-assemble. Our first tile set has the property that each scaled shape self-assembles via an asymptotically ``Kolmogorov-optimum'' temperature sequence but the scaling factor grows with the size of the shape being assembled. In contrast, our second tile set assembles each scaled shape via a temperature sequence whose length is proportional to the number of points in the shape but the scaling factor is a constant independent of the shape being assembled. We then show that there is no constant-size tile set that can uniquely assemble an arbitrary (non-scaled, connected) shape in the multiple temperature model, i.e., the scaling is necessary for self-assembly. This answers an open question of Kao and Schweller (\emph{Reducing Tile Complexity for Self-Assembly Through Temperature Programming}, SODA 2006), who asked whether such a tile set existed.
\end{abstract}

\clearpage

\section{Introduction}
Self-assembly is a process by which a small number of fundamental components automatically coalesce to form a target
structure. In 1998, Winfree \cite{Winf98} introduced the abstract Tile Assembly Model (aTAM) as an over-simplified discrete mathematical model of the DNA tile self-assembly pioneered by Seeman \cite{Seem82}. The aTAM is an ``effectivization'' of classical Wang tiling \cite{Wang61,Wang63} in which the fundamental components are un-rotatable, but translatable square ``tile types'' whose sides are labeled with glue ``colors'' and ``strengths.'' Two tiles that are placed next to each other \emph{interact} if the glue colors on their abutting sides
match, and they \emph{bind} if the strength on their abutting sides matches with total strength at least a certain ambient ``temperature.'' Extensive refinements of the aTAM were given by Rothemund and Winfree in \cite{RotWin00,Roth01}.

Despite its deliberate over-simplification, the aTAM is a computationally expressive model in the sense that Winfree \cite{Winf98} proved it is Turing-universal in two (or more) spatial dimensions. This suggests that it is possible, in principle, to algorithmically direct the process of self-assembly. The aTAM has also been studied from the perspective of computational complexity theory. A problem that has received substantial attention is that of finding ``small'' tile sets that assemble $N \times N$ squares in the aTAM. For instance, Adleman, Cheng, Goel, and Huang \cite{AdChGoHu01} proved that $N \times N$ squares self-assemble with $O\left(\frac{\log N} {\log\log N}\right)$ distinct tile types, matching the Kolmogorov-dictated lower bound that was established in \cite{RotWin00}. The more general problem of the self-assembly of arbitrary shapes in the aTAM has also been considered. Most notably, Soloveichik and Winfree \cite{SolWin07} discovered a beautiful connection between the the Kolmogorov complexity of an arbitrary scaled shape and the minimum number of tile types required to assemble it.

In addition to being an elegant and powerful theoretical tool, there is also experimental justification for the aTAM. For example, using DNA double-crossover molecules to construct tiles only a few nanometers long, Rothemund, Papadakis and Winfree \cite{RoPaWi04} implemented the molecular self-assembly of the well-known fractal structure called the \emph{discrete Sierpinski triangle} with low enough error rates to achieve correct placement of 100 to 200 tiles. Moreover, Barish, Schulman, Rothemund and Winfree \cite{BarSchRotWin09} have recently used Rothemund's DNA origami \cite{RotOrigami06} as a seed structure to which subsequent ``computation'' DNA tiles can attach and assemble computationally interesting patterns with error rates less than $.2\%$! Note that this technique, although robust, is not general-purpose in the sense that all of the information about the to-be-assembled shape (or pattern) is encoded into the DNA tiles and origami seed.

In fact, a central problem in algorithmic self-assembly is that of providing input to a tile assembly system (e.g., the size of a square, the description of a shape, etc.). In real-world laboratory implementations, as well as theoretical constructions, input to a tile system in the aTAM is provided via a (possibly large) collection of ``hard-coded'' seed tile types \cite{RotWin00,SolWin07,AdChGoHu01,BarSchRotWin09}. Unfortunately in practice, it is more expensive to manufacture many different types of tiles, as opposed to creating several copies of each tile type. This suggests that it might be advantageous to be able to provide input to a tile system without having to resort to hard-coding the input into a large number its own tiles. As a result, several natural generalizations of the aTAM have been developed in an attempt to model various types of alternative input delivery mechanisms.

One such model is the \emph{staged self-assembly} model \cite{Reif99localparallel,DDFIRSS07}, in which several intermediate structures are allowed to assemble in different test tubes before they are all mixed together to obtain the target structure. Demaine, Demaine, Fekete, Ishaque, Rafalin, Schweller, and Souvaine \cite{DDFIRSS07} proved that arbitrary shapes self-assemble with $O(1)$ tile types but with a corresponding increase in the number of stages and even (in some cases) an increase in the scale of the target shape. Note that, in the staged self-assembly model, the input to a tile system is implicitly encoded in the actions of the laboratory scientist and not in the tile types themselves.

Another means of providing input to a tile system is through the programming of the relative concentrations of its tile types. Becker, Rapaport, and R{\'e}mila \cite{BeckerRR06} proved that by appropriately setting the relative concentrations of tiles, squares, rectangles and diamonds can self-assemble in an expected sense with $O(1)$ tile types, but with a large (and undesirable) variance. Kao and Schweller \cite{KaoSchS08} improved the aforementioned result by showing that it is possible to program the relative concentrations of $O(1)$ tile types such that they will assemble into arbitrarily close approximations of $N \times N$ squares with high probability. Furthermore, Doty \cite{RSAES} recently showed that $N \times N$ squares self-assemble \emph{exactly} with high probability using $O(1)$ tile types.

The \emph{multiple temperature} model \cite{KS07,AGKS05} is a natural generalization of the aTAM, where the temperature of a tile system is dynamically adjusted by the experimenter as self-assembly proceeds. Aggarwal, Cheng, Goldwasser, Kao, and Schweller \cite{AGKS05} proved that the number of tile types required to assemble ``thin'' $k \times N$ rectangles can be reduced from $\Omega\left( \frac{N^{1/k}}{k}\right)$ (in the aTAM) to $\Omega\left( \frac{\log N}{\log \log N} \right)$ if the temperature is allowed to change but once. Subsequently, Kao and Schweller \cite{KS07} discovered a clever ``bit-flipping'' scheme capable of assembling any $N \times N$ square using $O(1)$ tile types and $\Theta(\log N)$ temperature changes. Note that the multiple temperature model has a similar flavor to that of the staged self-assembly model in the sense that the input to a tile system in both models can be encoded into a sequence of laboratory operations.

In all of the results mentioned in the previous three paragraphs, with the notable exception of \cite{DDFIRSS07}, attention was focused on the problem of reducing the number of distinct tile types needed for the assembly of certain \emph{restricted} classes of shapes such as diamonds, thin rectangles or squares. In this paper, we study the broader problem of reducing the number of tiles needed to assemble \emph{arbitrary} finite shapes in the multiple temperature model.

In particular, we exhibit two constant-size tile sets in which scaled-up versions of arbitrary shapes self-assemble. Our first tile set has the property that each scaled shape self-assembles via a temperature sequence whose length is proportional to the Kolmogorov complexity of the shape, but the scaling factor grows with the size of the shape being assembled. In contrast, our second tile set assembles each scaled shape via a temperature sequence whose length is proportional to the number of points in the shape but the scaling factor is a constant independent of the shape being assembled. Finally, we show that the scale factor in both of our constructions is necessary, i.e., that there is no constant-size tile set that can uniquely assemble an arbitrary shape in the multiple temperature model. This answers an open question of Kao and Schweller \cite{KS07}, who asked whether such a tile system existed.

The remainder of this paper is organized as follows. In Section 2, we review basic definitions and notation for both the abstract tile assembly model and the multiple temperature model. In Section 3, we exhibit two constant-size tile sets in which scaled-up versions of arbitrary shapes self-assemble in the multiple temperature model. In Section 4, we prove that there is no general-purpose tile set capable of the self-assembly of arbitrary shapes in the multiple temperature model. Section 5 contains concluding remarks and states an open question.

\section{Preliminaries}

We work in the $2$-dimensional discrete space $\Z^2$. Define the set
$U_2 = \{(0,1), (1,0), (0,-1), (-1,0)\}$ to be the set of all
\emph{unit vectors}, i.e., vectors of length 1 in $\mathbb{Z}^2$. We
write $[X]^2$ for the set of all $2$-element subsets of a set $X$.
All \emph{graphs} here are undirected graphs, i.e., ordered pairs $G
= (V, E)$, where $V$ is the set of \emph{vertices} and $E \subseteq
[V]^2$ is the set of \emph{edges}. A {\it grid graph} is a graph $G = (V, E)$ in which $V \subseteq \Z^2$ and
every edge $\{\vec{a}, \vec{b} \} \in E$ has the property that $\vec{a} -
\vec{b} \in U_2$.  The {\it full grid graph} on a set $V \subseteq \Z^2$ is the
graph $\fgg{V} = (V, E)$ in which $E$ contains {\it every} $\{\vec{a}, \vec{b}
\} \in [V]^2$ such that $\vec{a} - \vec{b} \in U_2$.

A \emph{shape} is a set $X \subseteq \mathbb{Z}^2$ such that $\fgg{X}$ is connected. In this paper, we consider scaled-up versions of finite shapes. Formally, if $X$ is a shape and $c \in \mathbb{N}$, then a $c$-\emph{scaling} of $X$ is defined as the set $X^c = \left\{ (x,y) \in \mathbb{Z}^2 \; \left| \; \left( \left\lfloor \frac{x}{c} \right\rfloor, \left\lfloor \frac{y}{c} \right\rfloor \right) \in X \right.\right\}$. Intuitively, $X^c$ is the shape obtained by replacing each point in $X$ with a $c \times c$ block of points. We refer to the natural number $c$ as the \emph{scaling factor} or \emph{resolution loss}. Note that scaled shapes have been studied extensively in the context of a variety of self-assembly systems \cite{SolWin07,DDFIRSS07,WinBek03,ChenGoel04}.

\subsection{The Abstract Tile Assembly Model}
We now give a brief and intuitive sketch of the aTAM that is adequate for reading this paper. More formal details and discussion may be found in \cite{Winf98,RotWin00,Roth01,jSSADST}.

Intuitively, a tile type $t$ is a unit square that can be
translated, but not rotated, having a well-defined ``side
$\vec{u}$'' for each $\vec{u} \in U_2$. Each side $\vec{u}$ of $t$
has a ``glue'' of ``color'' $\textmd{col}_t(\vec{u})$ -- a string
over some fixed alphabet $\Sigma$ -- and ``strength''
$\textmd{str}_t(\vec{u})$ -- a nonnegative integer -- specified by its type
$t$. Two tiles $t$ and $t'$ that are placed at the points $\vec{a}$
and $\vec{a}+\vec{u}$ respectively, \emph{bind} with \emph{strength}
$\textmd{str}_t\left(\vec{u}\right)$ if and only if
$\left(\textmd{col}_t\left(\vec{u}\right),\textmd{str}_t\left(\vec{u}\right)\right)
=
\left(\textmd{col}_{t'}\left(-\vec{u}\right),\textmd{str}_{t'}\left(-\vec{u}\right)\right)$.

Given a set $T$ of tile types, an \emph{assembly} is a partial
function $\pfunc{\alpha}{\Z^2}{T}$, with points $\vec{x}\in\Z^2$ at
which $\alpha(\vec{x})$ is undefined interpreted to be empty space,
so that $\dom \alpha$ is the set of points with tiles. An
assembly is $\tau$-\emph{stable}, where $\tau \in \mathbb{N}$, if it
cannot be broken up into smaller assemblies without breaking bonds
of total strength at least $\tau$. For an assembly $\alpha$, each $\vec{m} \in \mathbb{Z}^2$, and each $\vec{u} \in U_2$, $\text{str}_{\alpha}(\vec{m},\vec{u}) = \text{str}_{\alpha(\vec{m})}(\vec{u})\cdot \bval{\textmd{col}_{\alpha(\vec{m})}(\vec{u}) = \textmd{col}_{\alpha(\vec{m}+\vec{u})}(-\vec{u}) \textmd{ and } \textmd{str}_{\alpha(\vec{m})}(\vec{u}) = \textmd{str}_{\alpha(\vec{m}+\vec{u})}(-\vec{u})}$ where $\bval{\phi}$ is the {\it Boolean} value of the statement
$\phi$ (the Boolean value on the right is $0$ if $\{\vec{m},\vec{m}+\vec{u}\} \nsubseteq \dom{\alpha}$). The $\tau$-\emph{frontier} of an assembly $\alpha$, written as $\partial^\tau \alpha$ is the set of all points to which a tile can be $\tau$-stably added to $\alpha$.

Self-assembly begins with a \emph{seed assembly} $\sigma$ (typically
assumed to be finite and $\tau$-stable) and
proceeds asynchronously and nondeterministically, with tiles
absorbing one at a time to the existing assembly in any manner that
preserves stability at all times. All of the tile assembly systems in this paper are assumed to have a single seed tile placed at the origin.

A \emph{tile assembly system} (\emph{TAS}) is an ordered triple
$\mathcal{T} = (T, \sigma, \tau)$, where $T$ is a finite set of tile
types, $\sigma$ is a seed assembly with finite domain, and $\tau$ is
the temperature. An \emph{assembly sequence} in a TAS $\mathcal{T} = (T, \sigma, \tau)$ is a (possibly infinite) sequence $\vec{\alpha} = \left( \alpha_i \mid 0 \leq i < l\right)$ of assemblies in which $\alpha_0 = \sigma$ and each $\alpha_{i+1}$ is obtained from $\alpha_i$ by the ``$\tau$-stable'' addition of a single tile. Let $\vec{\alpha} = \left( \alpha_0, \alpha_1, \ldots\right)$ and $\vec{\alpha}' = \left( \alpha'_0, \alpha'_1, \ldots\right)$ be assembly sequences. We say that $\vec{\alpha}$ is a \emph{prefix} of $\vec{\alpha}'$, written as $\vec{\alpha} \sqsubseteq \vec{\alpha}'$, if there exists $j$ such that for every $0 \leq i \leq j$, $\alpha_i = \alpha'_i$. An assembly $\alpha$ is \emph{terminal}, and we write $\alpha \in
\termasm{\mathcal{T}}$, if no tile can be stably added to it. We
write $\termasm{\mathcal{T}}$ for the \emph{set of all terminal assemblies of} $\mathcal{T}$. A TAS ${\mathcal T}$ \emph{uniquely produces an assembly}, if it has exactly one terminal
assembly i.e., $|\termasm{\mathcal{T}}| = 1$.

\subsection{The Multiple Temperature Model}
In the multiple temperature model, a tile assembly system is defined as an ordered triple ${\mathcal T} = \left(T, \sigma, \left\langle \tau_i \right\rangle_{i=0}^{k-1}\right)$, where the third component $\left\langle \tau_i \right\rangle_{i=0}^{k-1}$ is a sequence of non-negative integer temperatures. The number $k$ is the \emph{temperature complexity} of $\mathcal{T}$. The \emph{temperature range} of a tile assembly system is the largest temperature in its temperature sequence.

\begin{figure}[htp]
    \begin{center}
      \subfloat[Tile types]{\label{fig:bitfliptiletypes}\includegraphics[width=1.0in]{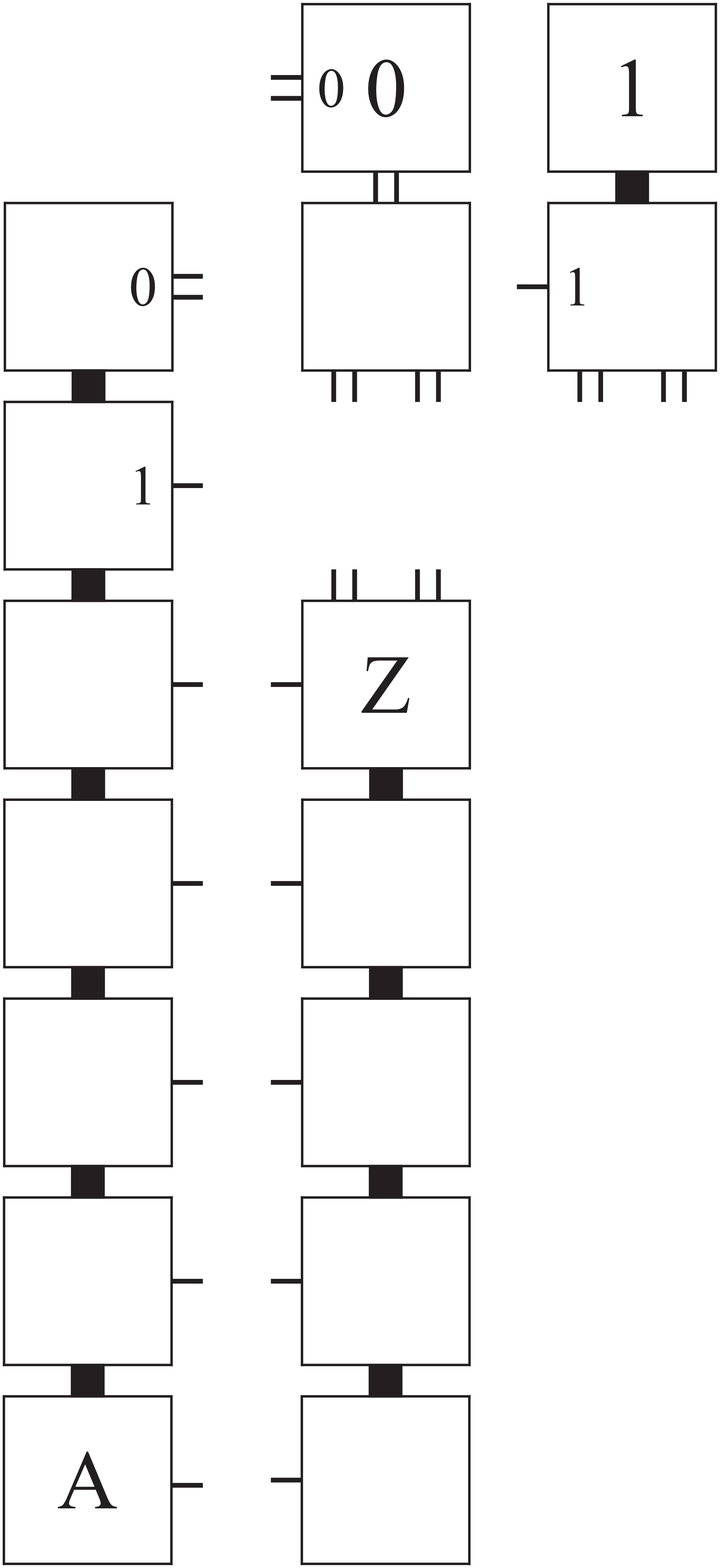}}
      \quad\quad\quad
      \subfloat[$\tau_0 = 2$]{\label{fig:bitflip0}\includegraphics[width=0.52in]{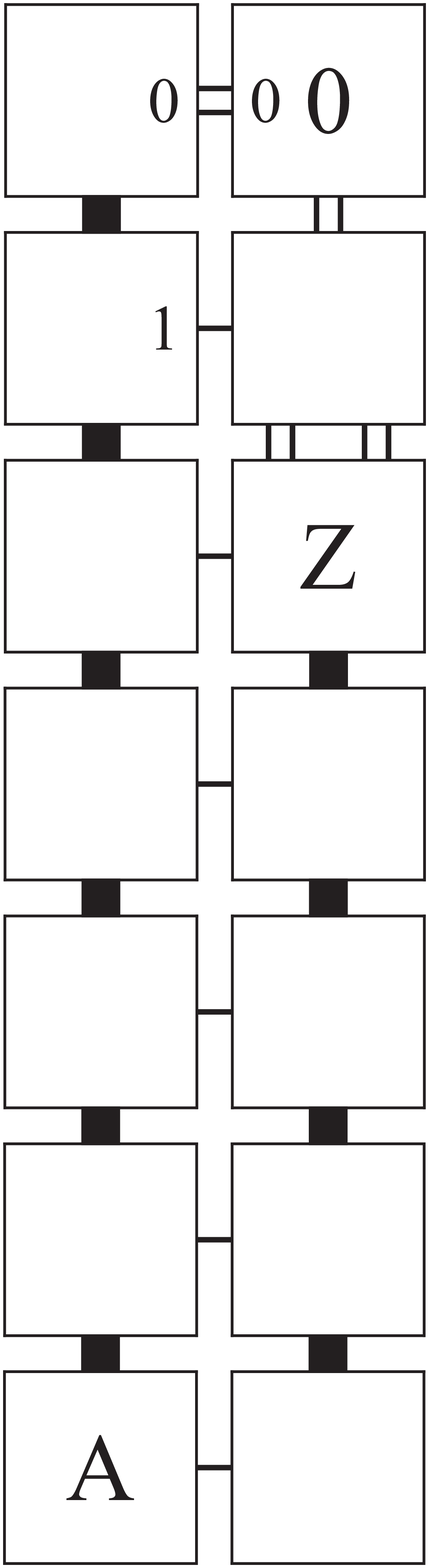}}
      \quad\quad\quad
      \subfloat[$\tau_1 = 5$]{\label{fig:bitflip1}\includegraphics[width=0.72in]{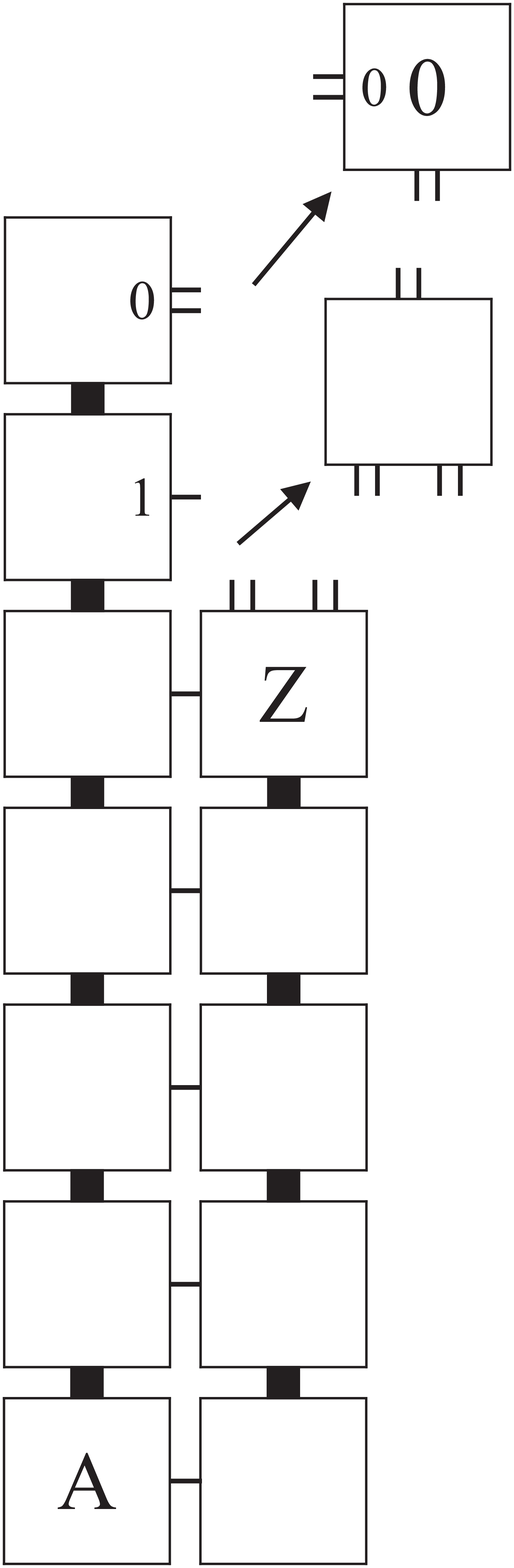}}
      \quad\quad\quad
      \subfloat[]{\label{fig:bitflip2}\includegraphics[width=0.67in]{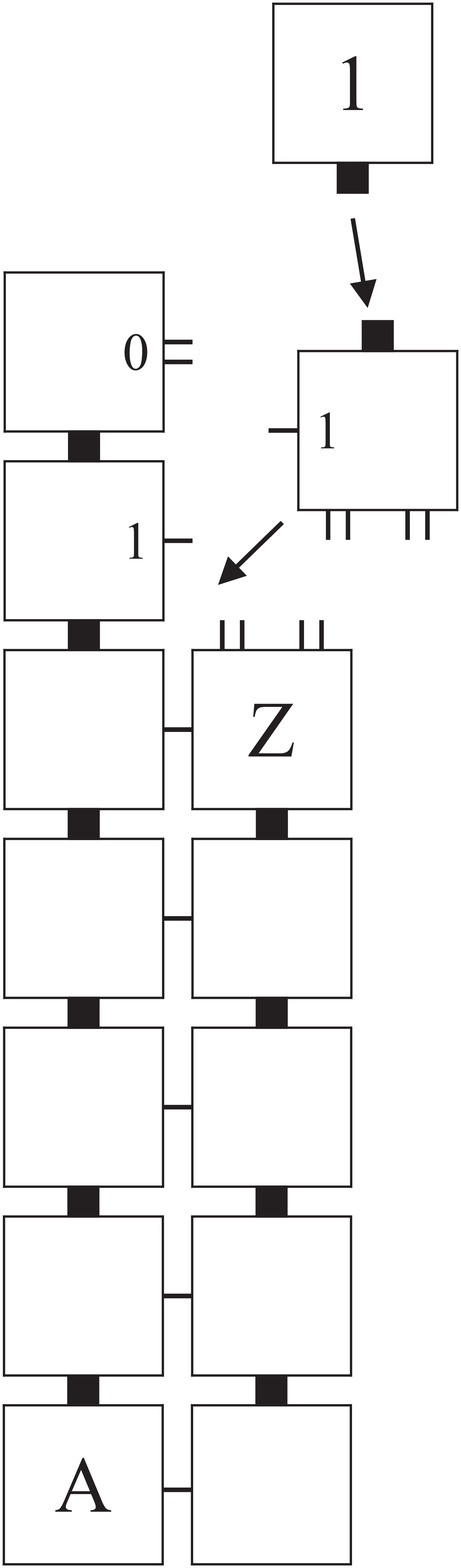}}
      \quad\quad\quad
      \subfloat[]{\label{fig:bitflip3}\includegraphics[width=0.52in]{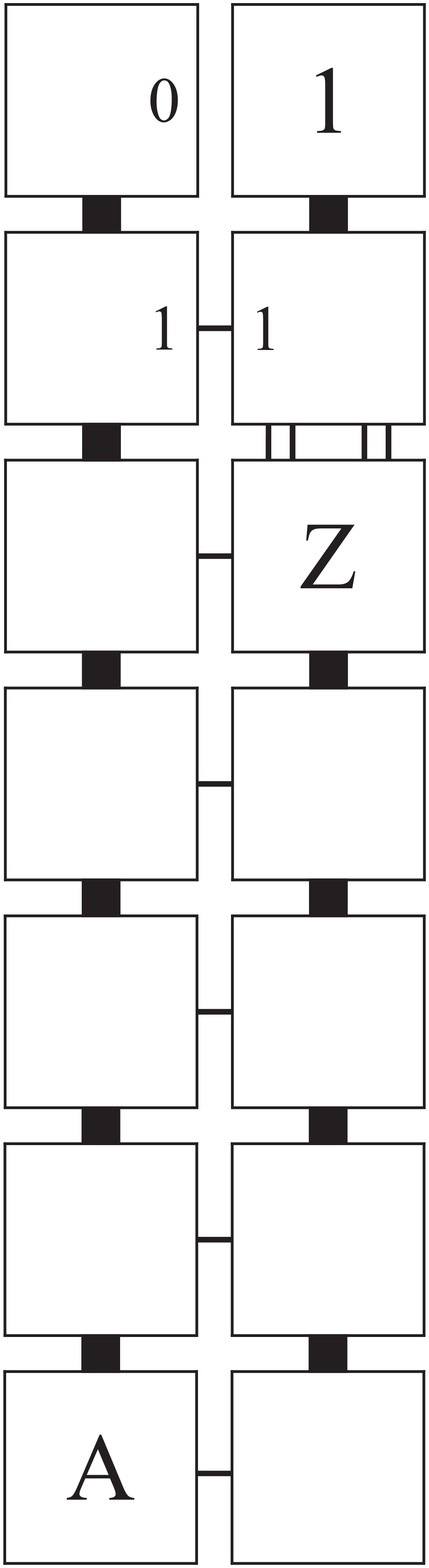}}
      \caption{\small \label{fig:bitflipexample} In this example, thick notches are strength 5, and thin notches are strength 1.}
      \vspace{-10pt}
    \end{center}
\end{figure}

Throughout the remainder of this section, let $\mathcal{T} = \left(T,\sigma,\left \langle \tau_i \right \rangle_{i=0}^{k-1} \right)$ be a multiple temperature tile assembly system. Intuitively, self-assembly in $\mathcal{T}$ is carried out in $k$ phases. In the first temperature phase, tiles are added to the existing assembly as they normally would be in the abstract model until a $\tau_0$-stable terminal assembly is reached. In phase two, tiles can accrete to the existing assembly if they can do so with at least strength $\tau_1$. Also, and at any time during the second temperature phase, if there is ever a cut of the assembly having a strength less than $\tau_1$, then all of the tiles on the side of the cut not containing the seed can be removed from the assembly. When a $\tau_1$-stable terminal assembly is reached in phase two, phase three begins and proceeds in a similar fashion. This process continues through the final temperature phase in which tiles are added or removed with respect to the temperature $\tau_{k-1}$ until reaching a $\tau_{k-1}$-stable terminal assembly. See Figure~\ref{fig:bitflipexample} for a specific example of the self-assembly of a bit-flip gadget \cite{KS07} in the multiple temperature model.

We define an \emph{assembly sequence} for the $i^{\textmd{th}}$ temperature phase of $\mathcal{T}$ as follows. If $i = 0$, then an assembly sequence for temperature phase $0$ of $\mathcal{T}$ is an assembly sequence $\vec{\alpha}$ of the tile assembly system $\mathcal{T}_0 = (T,\sigma,\tau_0)$. For $i > 0$, an assembly sequence for temperature phase $i$ of $\mathcal{T}$ is a (possibly infinite) sequence of assemblies $\vec{\alpha} = (\alpha_0, \alpha_1, \ldots)$ satisfying the following conditions.
\begin{enumerate}
    \item There exists $l$ such that $\alpha_l$ is $\tau_{i-1}$-stable, $\partial^{\tau_{i-1}}\alpha_l = \emptyset$ and $(\alpha_0,\alpha_1,\ldots,\alpha_l)$ is an assembly sequence for temperature phase $i-1$ of $\mathcal{T}$; and
    \item for all $j \geq l$, $\alpha_{j+1}$ is obtained from $\alpha_j$ by the $\tau_{i}$-stable addition of a single tile or the deletion of a cut of $\alpha_j$ - to which the seed tile does not belong - having strength less than $\tau_{i}$.
\end{enumerate}
An \emph{assembly sequence} in $\mathcal{T}$ is an assembly sequence for the $i^{\textmd{th}}$ temperature phase of $\mathcal{T}$ for some $i \in \mathbb{N}$. We say that an assembly sequence $\vec{\alpha}$ in $\mathcal{T}$ \emph{finishes every temperature phase} of $\mathcal{T}$ if $\vec{\alpha}$ is a finite assembly sequence for temperature stage $k-1$ of $\mathcal{T}$ and its final assembly, denoted $\alpha$, is $\tau_{k-1}$-stable and $\partial^{\tau_{k-1}}\alpha = \emptyset$. In this case, we call $\alpha$ a \emph{terminal assembly} and write $\alpha \in \termasm{\mathcal{T}}$.
If, for every assembly sequence $\vec{\alpha}$ in $\mathcal{T}$, there exists an assembly sequence $\vec{\alpha}'$ in $\mathcal{T}$ such that $\vec{\alpha}'$ finishes every temperature phase of $\mathcal{T}$, $\vec{\alpha} \sqsubseteq \vec{\alpha}'$, and $\termasm{\mathcal{T}} = \{ \alpha \}$, then $\mathcal{T}$ \emph{uniquely produces the assembly} $\alpha$. For a given shape $X$, we say that $\mathcal{T}$ \emph{uniquely produces the shape} $X$ if $\termasm{\mathcal{T}} = \{ \alpha \}$ and $\dom{\alpha} = X$.

\section{Self-Assembly of Arbitrary Scaled Shapes with $O(1)$ Tile Types}

In this section, we exhibit two constructions that are capable of building scaled-up versions of arbitrary shapes in the multiple temperature model. Both constructions reduce the tile complexity for the self-assembly of arbitrary scaled shapes from $\Theta\left(\frac{K(X)}{\log K(X)}\right)$ \cite{SolWin07} to $O(1)$, but with a corresponding increase in temperature complexity.
\begin{figure}[htp]
\centering
    \subfloat[][Partial seed block: west growing simulation of $U$ on $\pi$]{%
        \label{fig:firstconstruction0}%
        \includegraphics[width=3.50in]{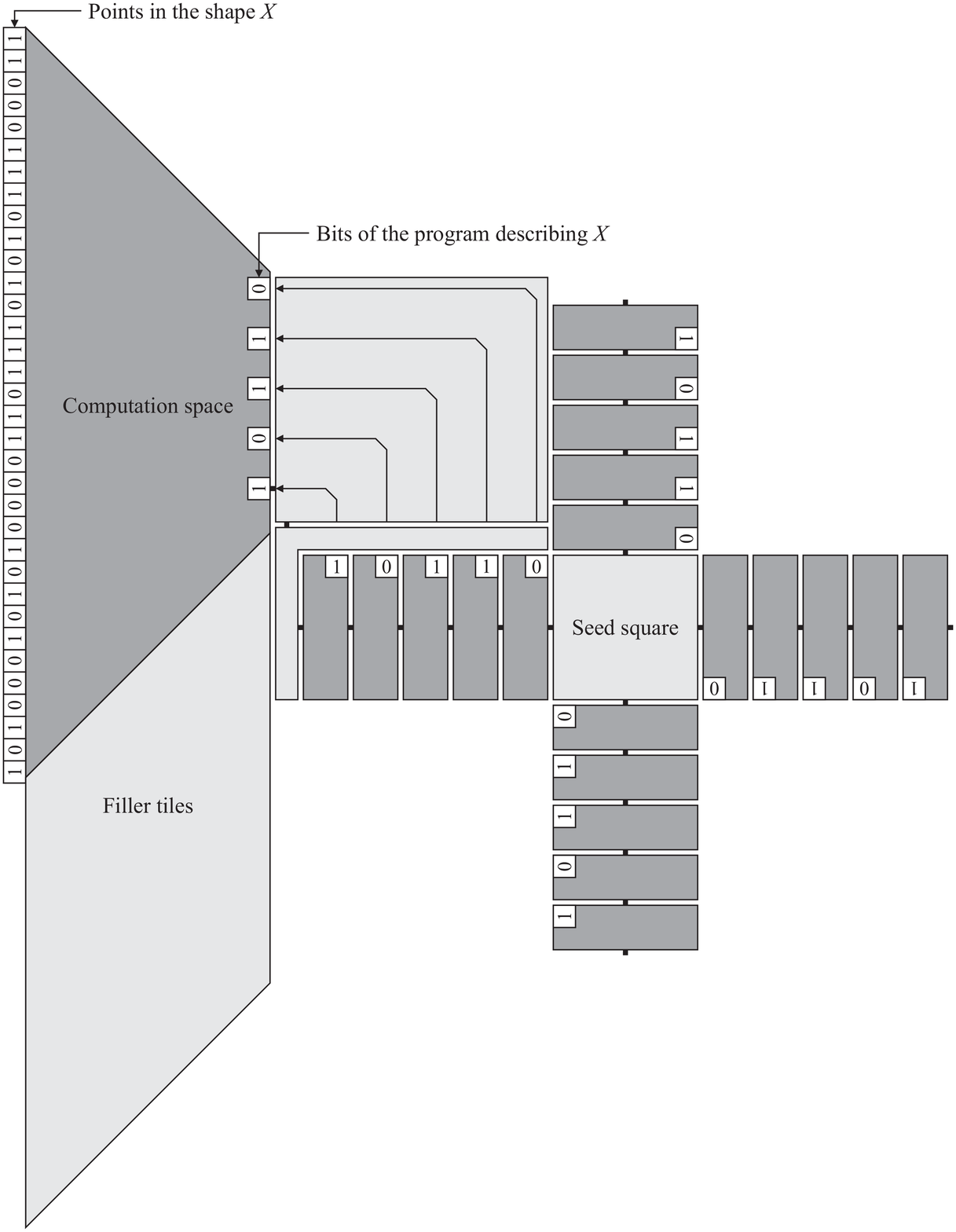}}%
        \hspace{30pt}%
    \subfloat[][A completed seed block]{%
        \label{fig:firstconstruction1}%
        \includegraphics[width=1.75in]{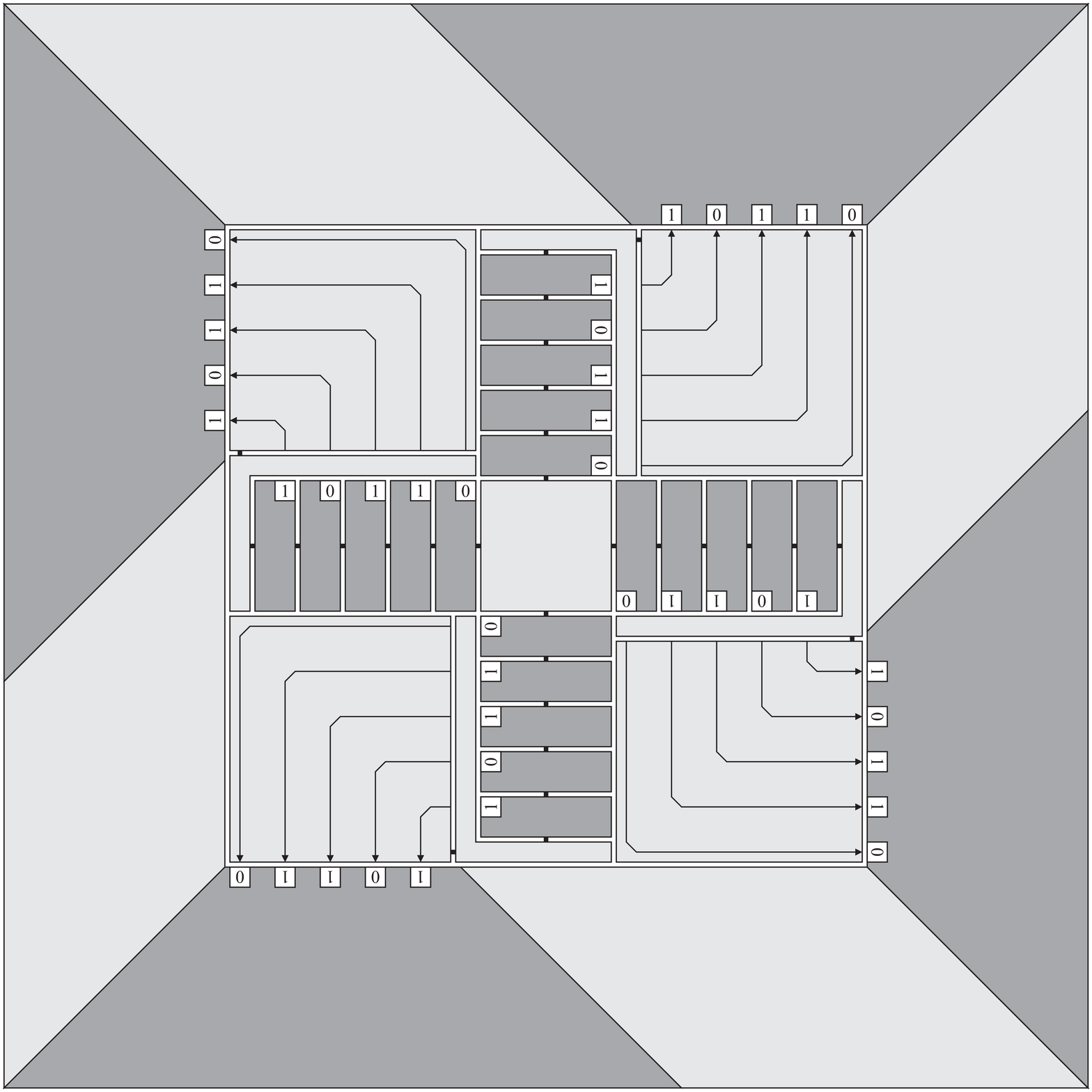}}\\
    \caption{\small In our first construction, the seed square assembles first. Then four identical rectangles simultaneously assemble off each side of the seed square via the corresponding temperature sequence defined in the proof of Theorem 3.1 in \cite{KS07}. After the self-assembly of the four rectangles, the rotated `L' structure initiates the assembly of a square in which the bits of $\pi$ are rotated up and to the left into the computation space. Finally, $U$ is simulated on $\pi$ in the computation space. After the simulation is done, the shape $X$ is encoded along the border of the seed block. The filler tiles are used to ensure that the seed block is a square.}%
        \label{fig:firstconstruction}%
    \vspace{-10pt}
\end{figure}
\subsection{Optimum Temperature Sequences but Unbounded Scaling Factors}
In our first construction, we simply combine a portion of the main construction of \cite{SolWin07} with the bit-flip gadget of \cite{KS07}. Fix some universal Turing machine $U$. The \emph{Kolmogorov complexity} of a shape $X$, denoted as $K(X)$, is the size of the smallest program $\pi$ that outputs an encoding of a list of all the points in $X$. In other words $K(X) = \min\{ |\pi| \mid U(\pi) = \langle S \rangle \}$. The reader is encouraged to consult \cite{LiVitanyiIntro} for a more detailed discussion of Kolmogorov complexity.

\begin{theorem}
\label{secondmaintheorem}
There exists a tile set $T$ with $|T| = O(1)$ such that, for every finite shape $X$, there exists $c \in \mathbb{N}$ and a temperature sequence $\left \langle \tau_i \right\rangle_{i=0}^{k-1}$ with $k = O(K(X))$ such that $\mathcal{T}^{X^c} = \left(T,\sigma,\left \langle \tau_i \right\rangle_{i=0}^{k-1}\right)$ uniquely produces $X^c$.
\end{theorem}

\begin{proof}
The basic idea is to combine the tile set (of Theorem 3.1) of \cite{KS07} that assembles an $11 \times 2m$ rectangle (whose top row encodes the bits of an arbitrary binary string of length $m$) with the portion of the tile set from \cite{SolWin07} that does not contain any ``seed-frame'' or ``un-packing'' tile types. Thus, given any program $\pi$ that, when run by $U$, outputs $X$ as a list of points, we can use bit-flip gadgets to encode a description of $\pi$. Then the main construction from \cite{SolWin07} can proceed normally at temperature $\tau_{k-1} = 2$. The only (minor) technicality is that we must add additional tile types to ensure that the seed block is properly assembled, which we illustrate in Figure~\ref{fig:firstconstruction}. Note that the size of this tile set is $O(1)$, i.e., it is independent of the shape being assembled, and the temperature complexity is $O(|\pi|)$.
\end{proof}

\begin{remark}
Theorem~\ref{secondmaintheorem} is tight. For instance, for an algorithmically random string $w = w_0w_1,\cdots w_{n-1}$, the shape $X(w) = \left\{ \left(x_0,y_0\right),\left(x_1,y_1\right),\ldots \left(x_{n-1},y_{n-1}\right)\right\}$ defined by $\left(x_0,y_0\right) = (0,0)$, and $\left(x_{i+1},y_{i+1}\right) = \left(x_{i+1},y_i+1\right)$ if $w_i = 0$, and $\left(x_{i}+1,y_{i+1}\right)$ if $w = 1$ (i.e., a path that goes right if $w_i = 0$ and up if $w_i = 1$), has $K(X(w)) \approx K(w) \geq n$.
\end{remark}

Note that in the construction for Theorem~\ref{secondmaintheorem}, the scaling factor can be quite large. Specifically, the scaling factor $c$ depends on the running time of $\pi$, whence $c = poly(time(\pi))$ \cite{SolWin07}. Also, the scaling factor in the above construction is further inflated (albeit by a constant factor) by the filler tiles in the assembly of the seed block. In a truly nano-scale setting, it is necessary to have a construction in which the scaling factor is always small, or better yet, bounded by a constant independent of the shape being assembled. We now show how to achieve this.

\subsection{Constant Scaling Factor but Long Temperature Sequences}
Recall that for any scaled finite shape $X^c$, each point in $X$ gets mapped to a $c \times c$ block of points in $X^c$. In our second construction, we will build a simple \emph{square gadget} that will be responsible for the assembly of each $c \times c$ block in $X^c$. As a result, the scaling factor $c$ in our second construction will depend entirely on the size of the square gadgets.
\begin{figure}[htp]
    \begin{center}
        \subfloat[Go ``straight'']{\label{fig:square_continue}\includegraphics[width=0.70in]{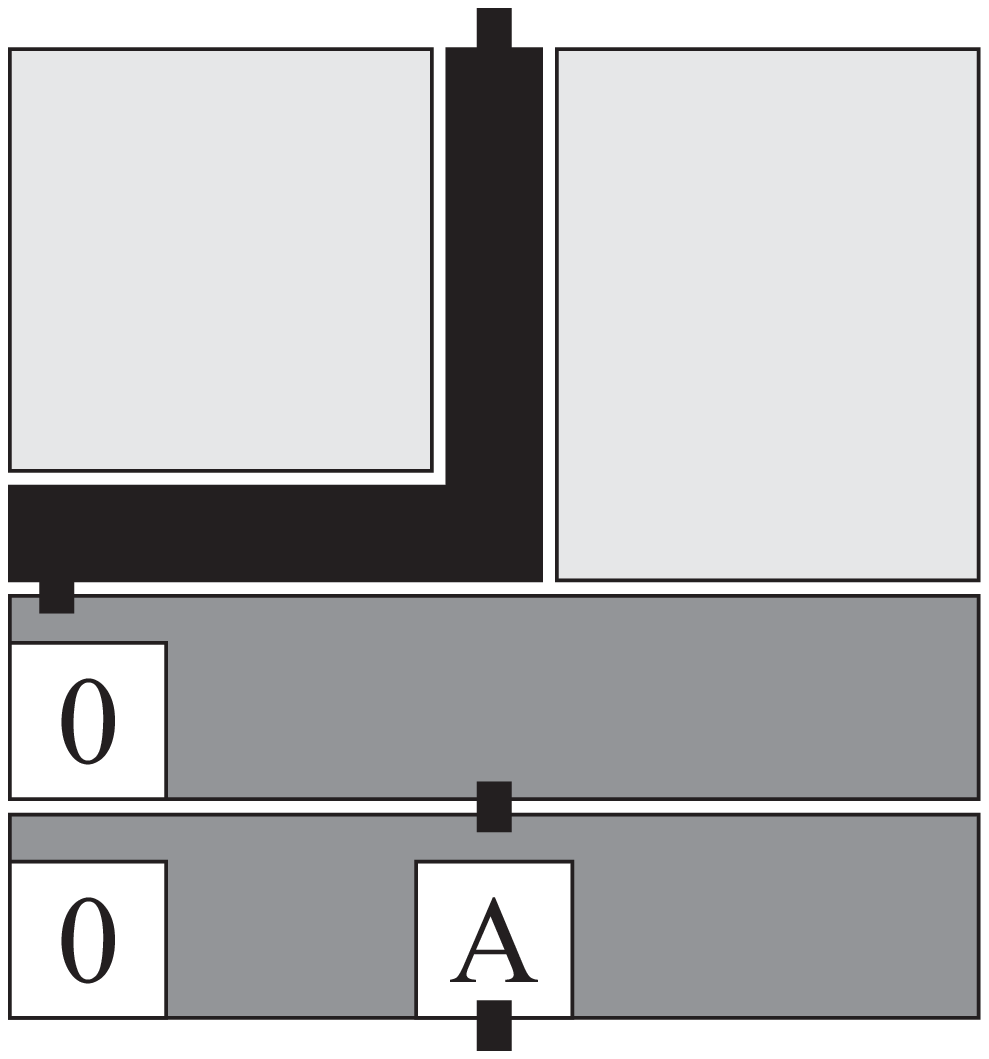}}
        \hspace{20pt}
        \subfloat[Turn left]{\label{fig:square_left}\includegraphics[width=0.70in]{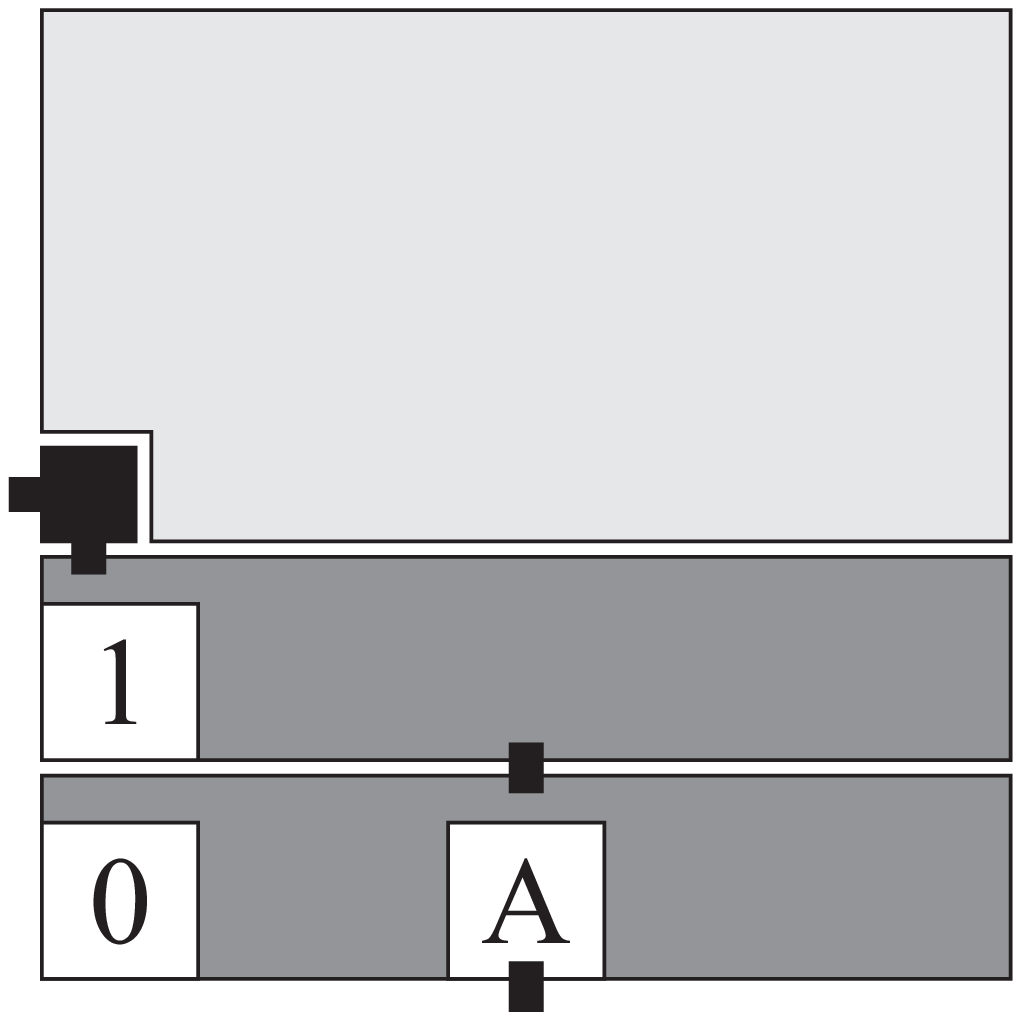}}
        \hspace{20pt}
        \subfloat[Turn right]{\label{fig:square_right}\includegraphics[width=0.70in]{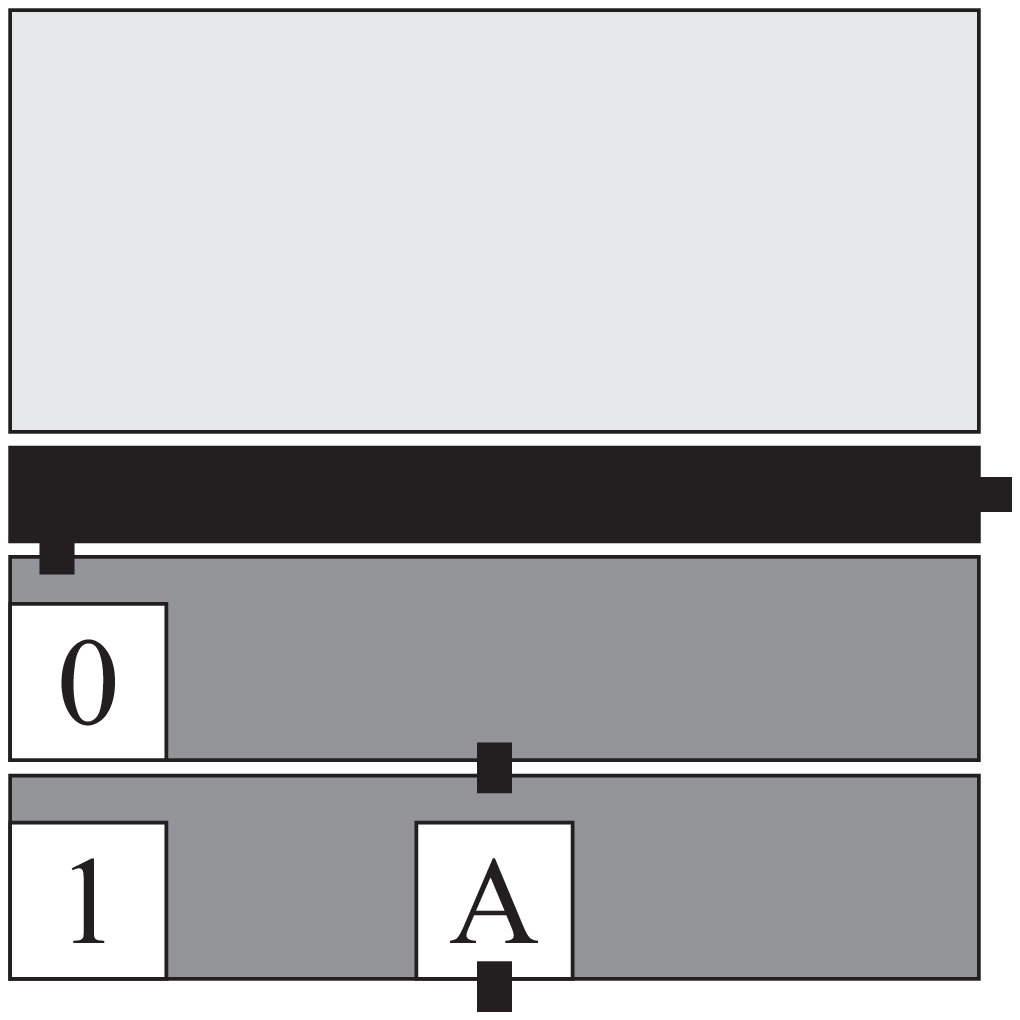}}
        \hspace{20pt}
        \subfloat[Stop]{\label{fig:square_stop}\includegraphics[width=0.70in]{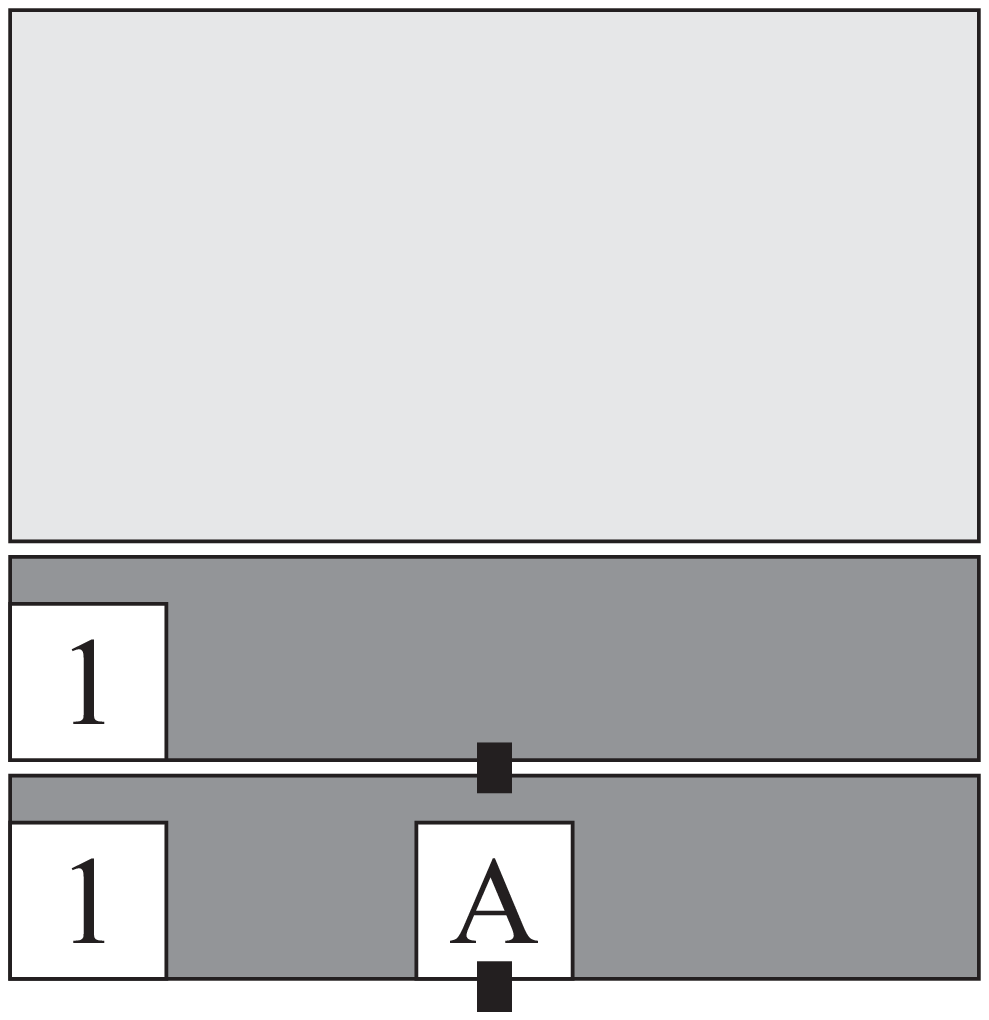}}
        \caption{\small \label{fig:square_gadgets} Overview of square gadgets: each square gadget consists of two bit-flip gadgets. The second (a.k.a. upper) bit-flip gadget remembers the value of the first gadget and thus can initiate the correct change in direction. The little black notches on the borders of the first three square gadgets initiate the growth of another (appropriately-rotated) square gadget.}
    \end{center}
    \vspace{-10pt}
\end{figure}

\begin{figure}[htp]%
\centering
    \subfloat[][$\langle 4 \rangle$]{%
        \label{fig:square0}%
        \includegraphics[width=3.0in]{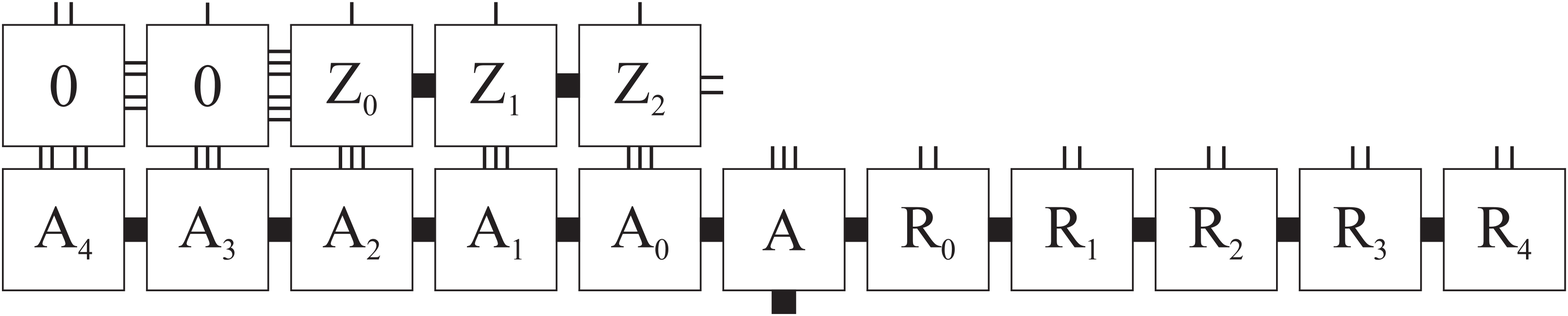}}%
        \hspace{15pt}%
    \subfloat[][$\langle 4, 3 \rangle$]{%
        \label{fig:square1}%
        \includegraphics[width=3.0in]{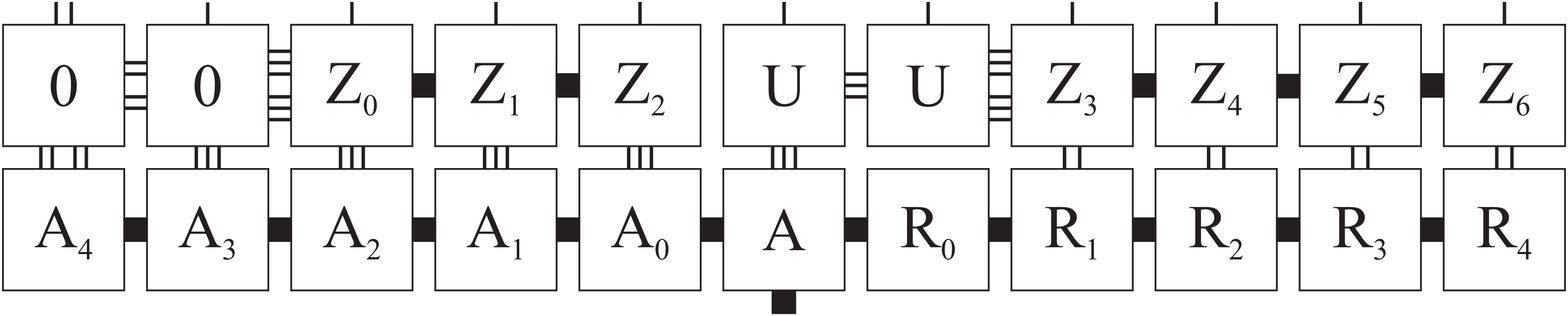}}\\
        \vspace{10pt}
    \subfloat[][$\langle 4, 3, 8 \rangle$]{%
        \label{fig:square2}%
        \includegraphics[width=3.0in]{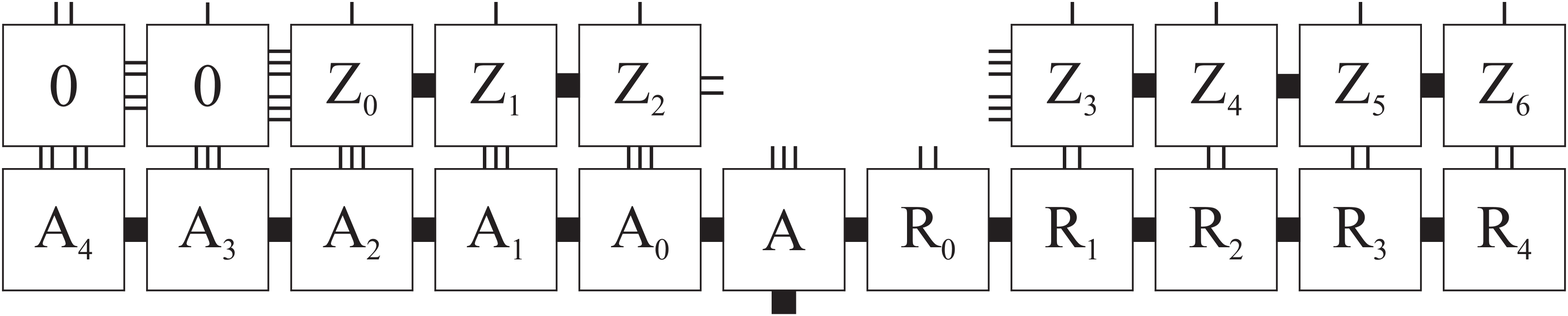}}%
        \hspace{15pt}%
    \subfloat[][$\langle 4, 3, 8 \rangle$; Note that $B_4$ binds cooperatively and thus ``remembers'' the bit of the first bit-flip gadget.]{%
        \label{fig:square3}%
        \includegraphics[width=3.0in]{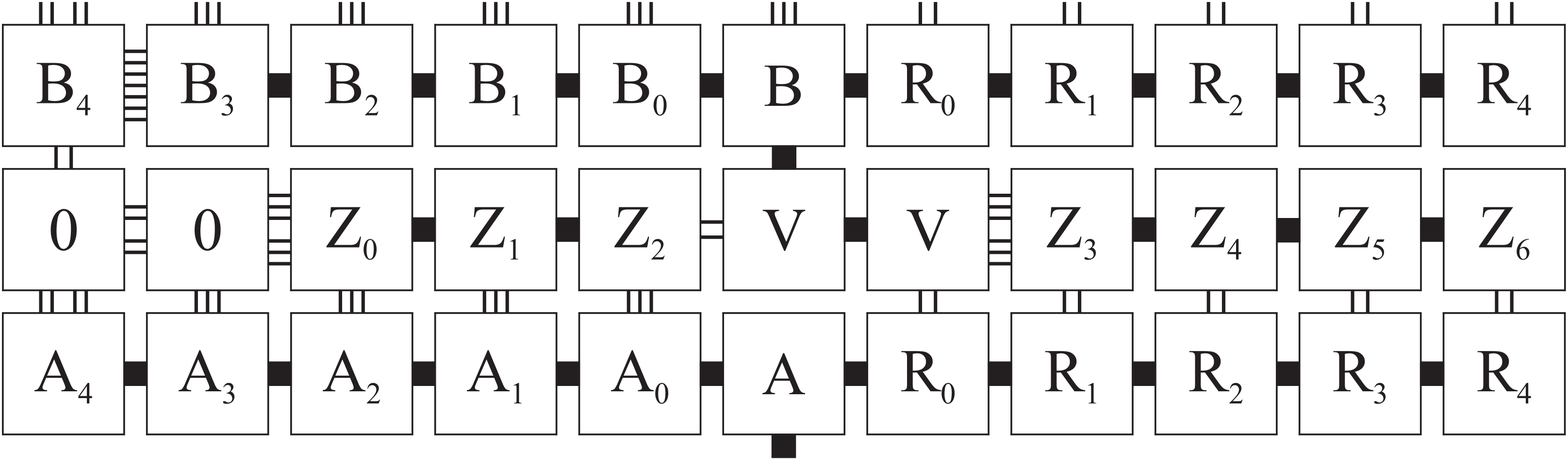}}\\
    \subfloat[][$\langle 4, 3, 8, 4 \rangle$]{%
        \label{fig:square4}%
        \includegraphics[width=3.0in]{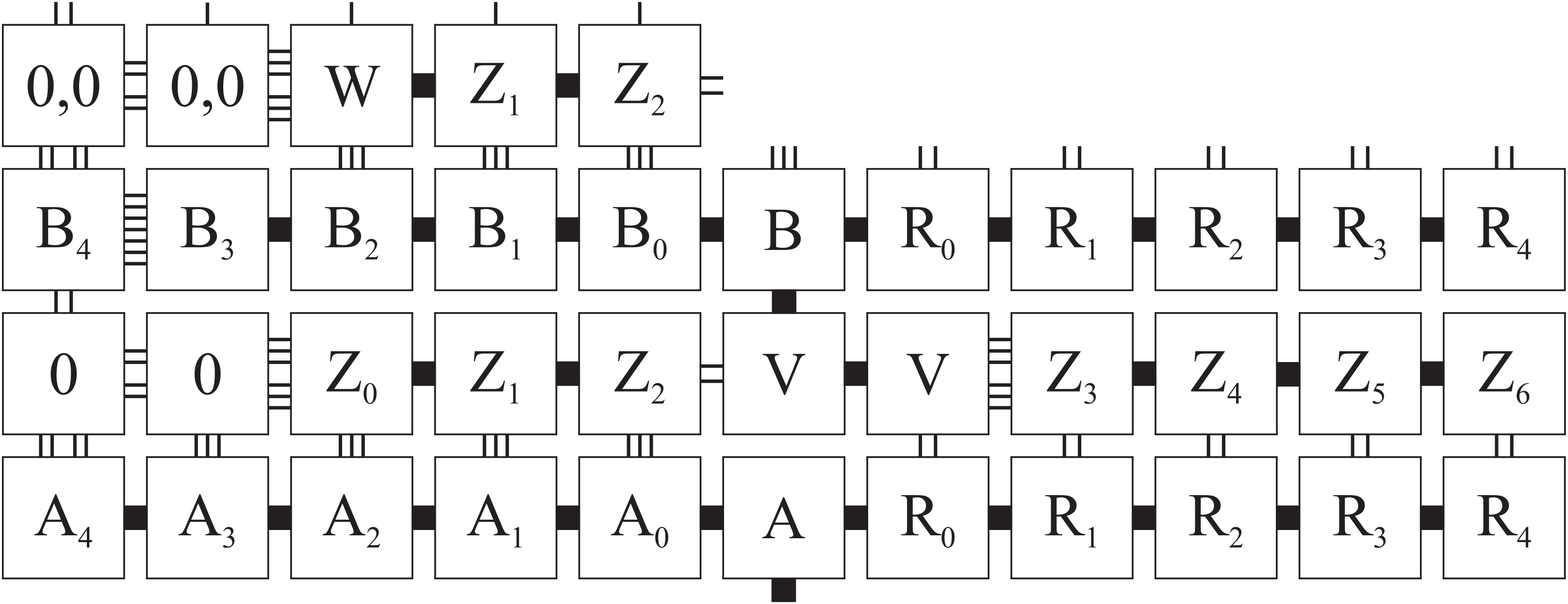}}%
        \hspace{15pt}%
    \subfloat[][$\langle 4, 3, 8, 4, 9 \rangle$; the tile in the upper-left corner has been flipped from 0 to 1 while still remembering the value of the previous bit-flip gadget.]{%
        \label{fig:square5}%
        \includegraphics[width=3.0in]{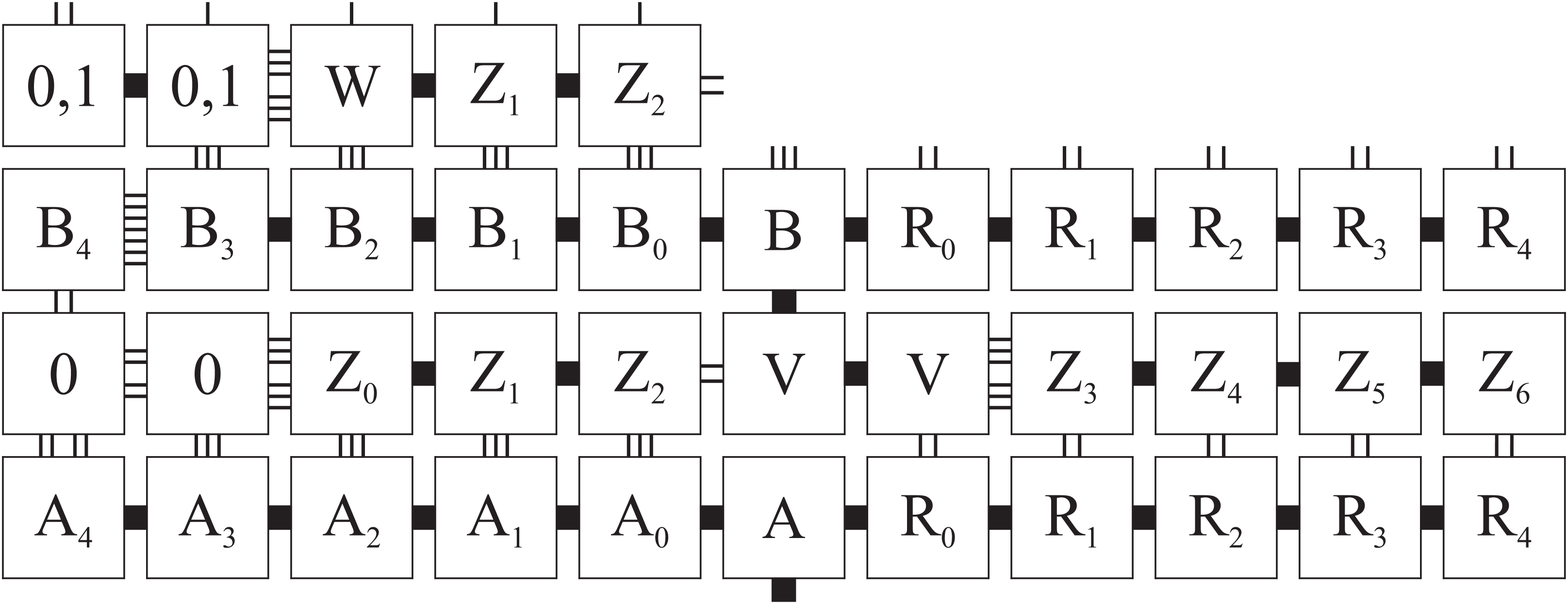}}\\
        \vspace{10pt}
    \subfloat[][$\langle 4, 3, 8, 4, 9, 3 \rangle$]{%
        \label{fig:square6}%
        \includegraphics[width=3.0in]{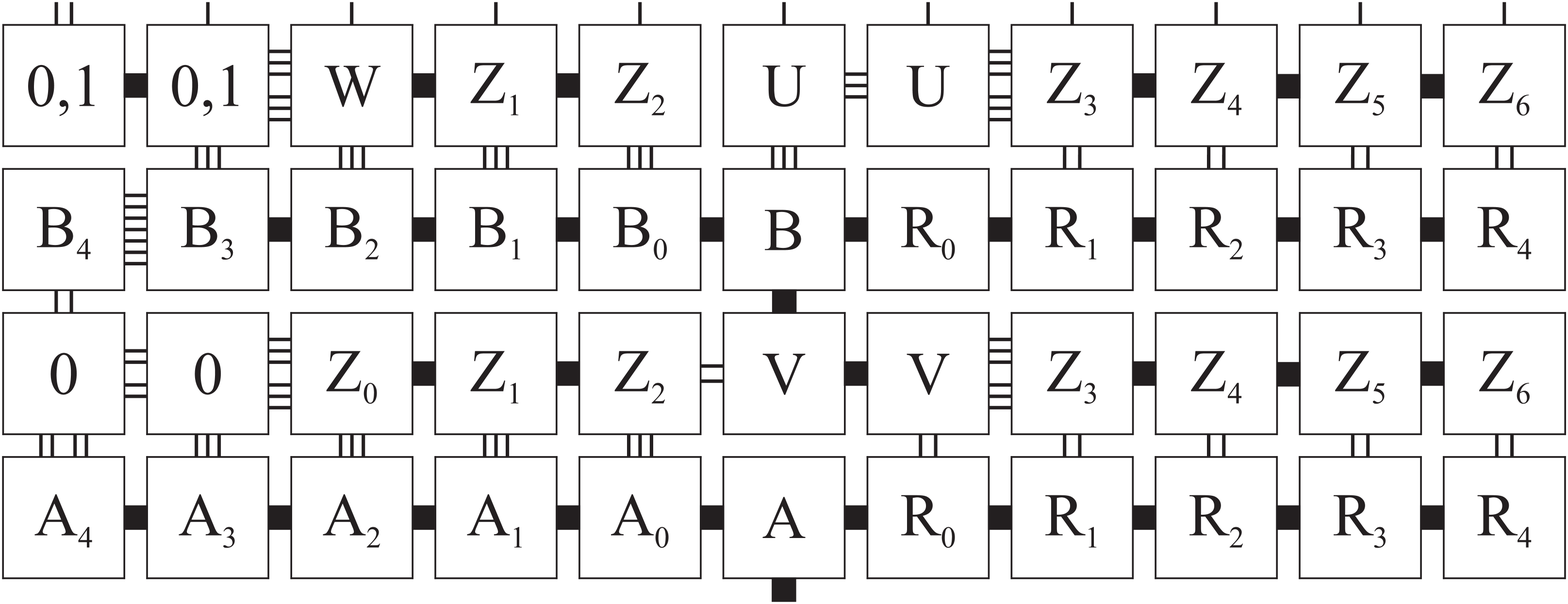}}%
        \hspace{15pt}%
    \subfloat[][$\langle 4, 3, 8, 4, 9, 3, 8 \rangle$]{%
        \label{fig:square7}%
        \includegraphics[width=3.0in]{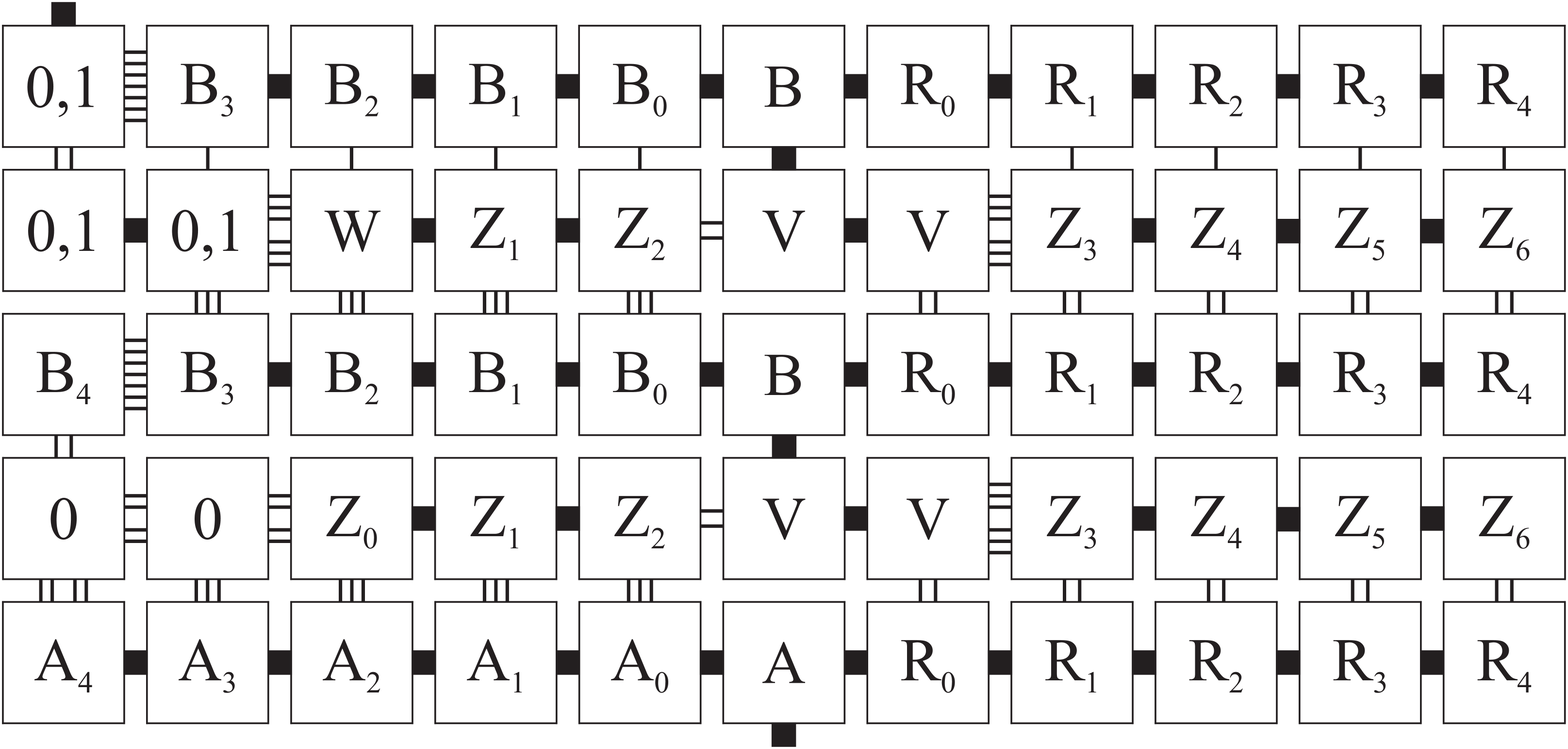}}%
        \hspace{15pt}%
    \caption{\small \label{fig:example} An example of programming a square gadget to ``turn left.'' We do this by setting the first bit-flip gadget to 0. Then we set the second bit-flip gadget to 1. Note that the latter bit-flip gadget remembers the value of the former bit-flip gadget.}%
\end{figure}

\begin{figure}[htp]
    \begin{center}
    \includegraphics[width=3.0in]{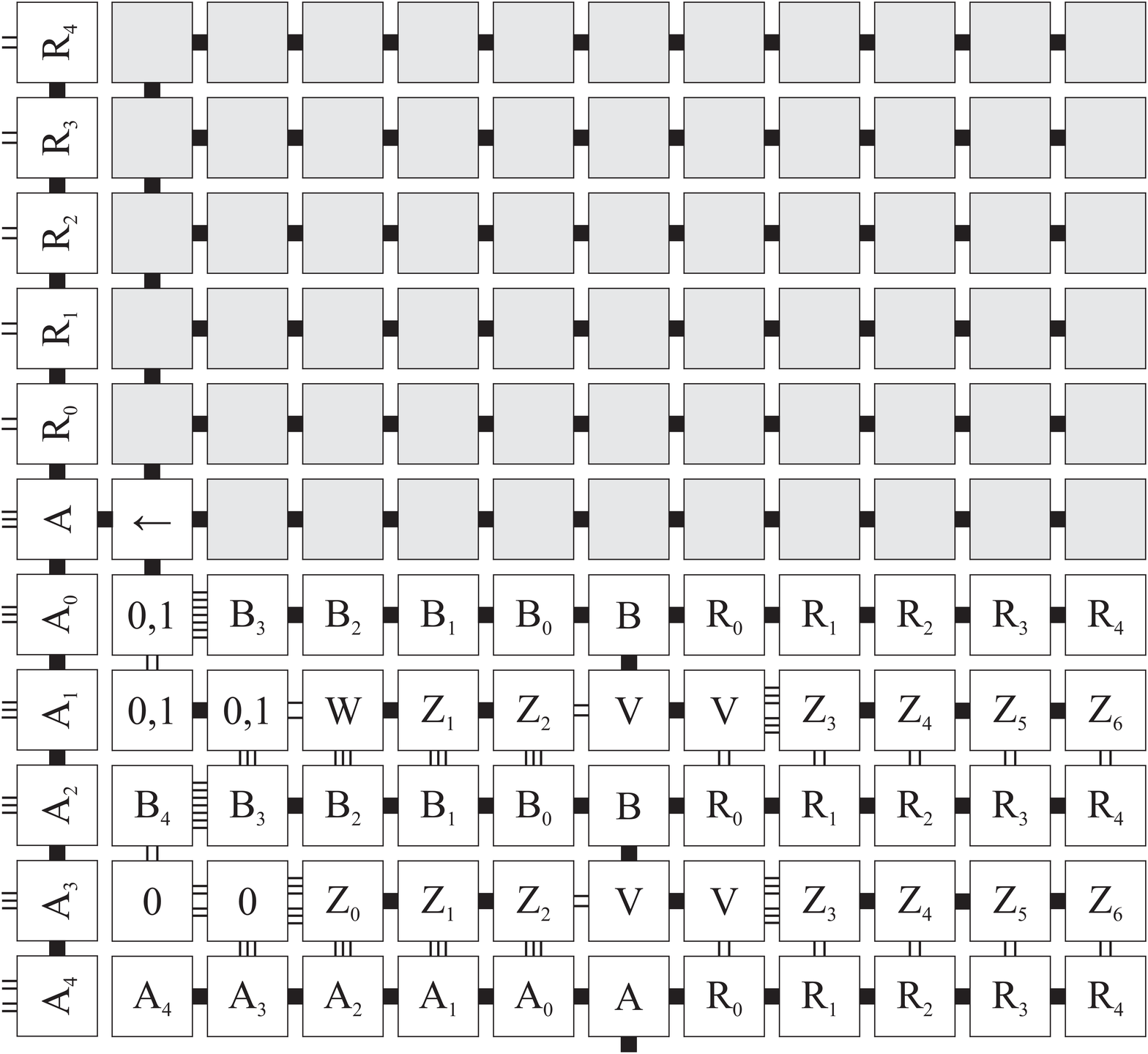}
    \caption{\small \label{fig:square8} $\langle 4, 3, 8, 4, 9, 3, 8 \rangle$; The final result of Figure~\ref{fig:example}. Note the special tile (labeled with `$\leftarrow$') along the left border of the completed square gadget initiated the growth of (the first column of) an appropriately rotated square gadget.}
    \end{center}
\end{figure}

Intuitively, each square gadget consists of two logical components: a lower and an upper half. The lower half of a square gadget is the concatenation of two bit-flip gadgets such that the second bit-flip gadget ``remembers'' the value of the first. The upper half of each square gadget then places a special output tile along the left, top or right side of the square depending on the values of the bit-flip gadgets in the lower half. Finally, the special output tile initiates the growth of another (appropriately-rotated) square gadget and the process is repeated for every point in the shape. Figure~\ref{fig:square_gadgets} gives an intuitive overview of the four canonical square gadgets. A more detailed example of the self-assembly of a square gadget is shown in Figures~\ref{fig:example} and~\ref{fig:square8} in the technical appendix.

\begin{theorem}
\label{assemble_ham_cycle}
There exists a tile set $T$ with $|T| = O(1)$, such that, for every finite shape $X$, if there is a Hamiltonian path $C$ in $\fgg{X}$, then there exists a temperature sequence $\left \langle \tau_i \right\rangle_{i=0}^{k-1}$ with $k = O(|X|)$, such that $\mathcal{T}^{X^{11}} = \left(T,\sigma,\left \langle \tau_i \right\rangle_{i=0}^{k-1}\right)$ uniquely produces $X^{11}$.
\end{theorem}

\begin{figure}[htp]
    \begin{center}
        \subfloat[Example shape $X$]{\label{fig:secondconstruction0}\includegraphics[width=0.65in]{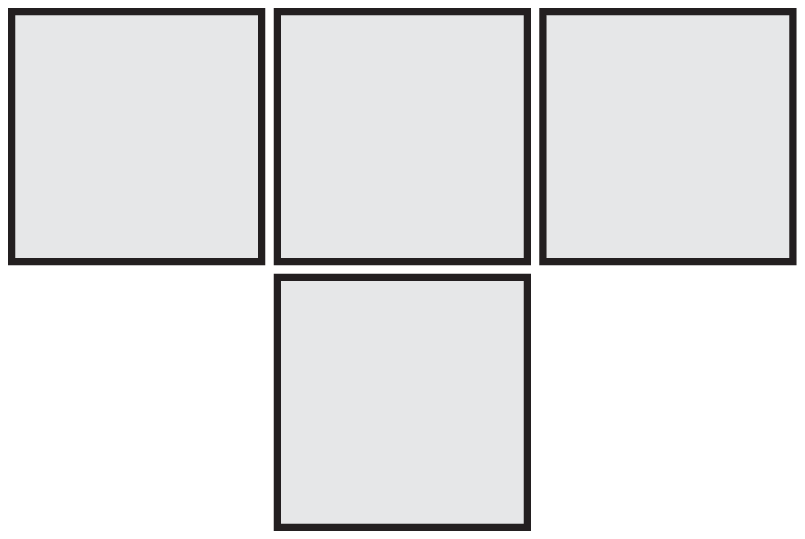}}
        \hspace{30pt}
        \subfloat[Hamiltonian path of $\fgg{X^2}$]{\label{fig:secondconstruction1}\includegraphics[width=1.10in]{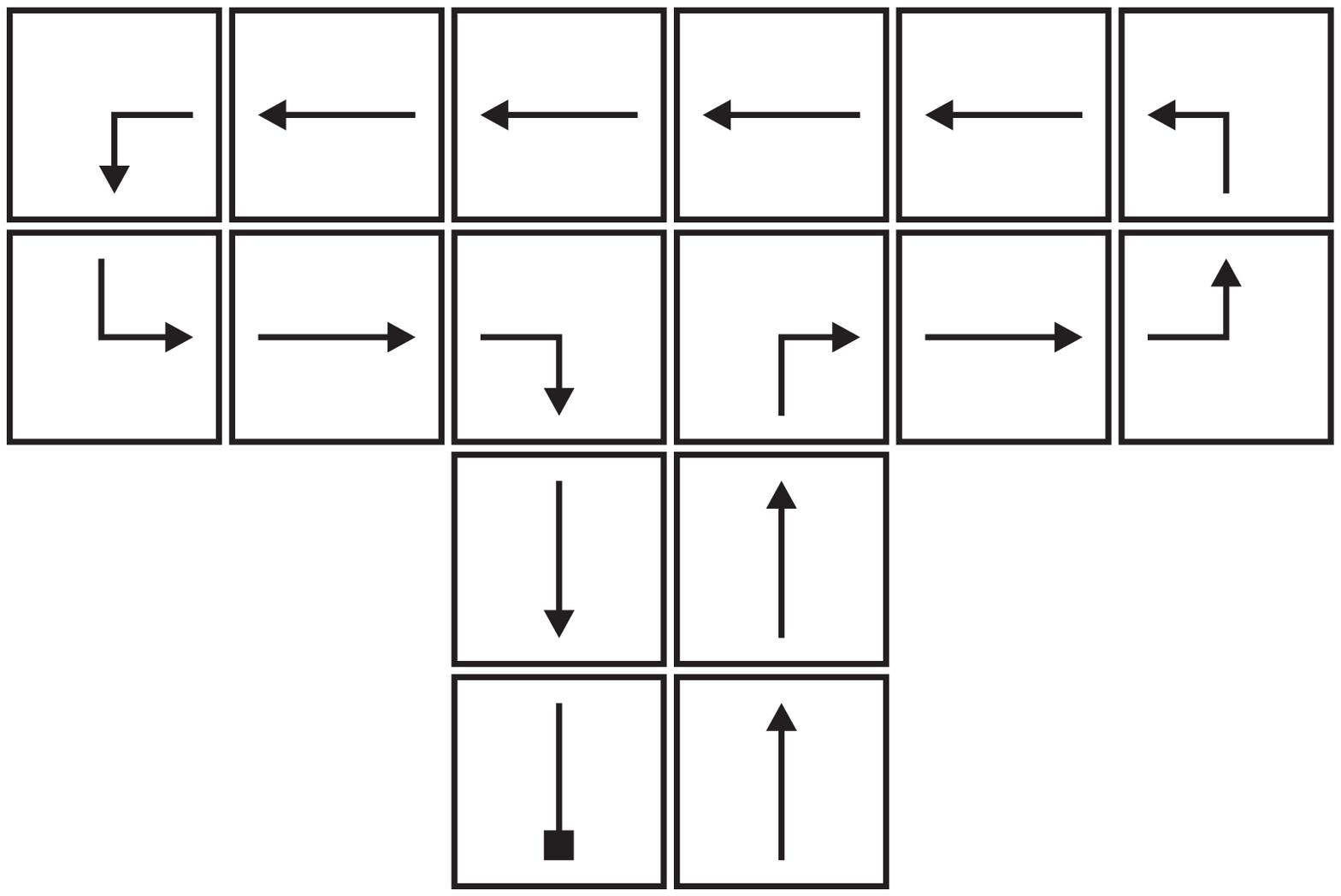}}
        \hspace{30pt}
        \subfloat[Self-assembly of $X^{22}$. The `S' tile is the single seed tile. Notice that the first three square gadgets have the same type of direction, while the fourth square gadget is rotated $90^\circ$ clockwise.]{\label{fig:secondconstruction2}\includegraphics[width=3.00in]{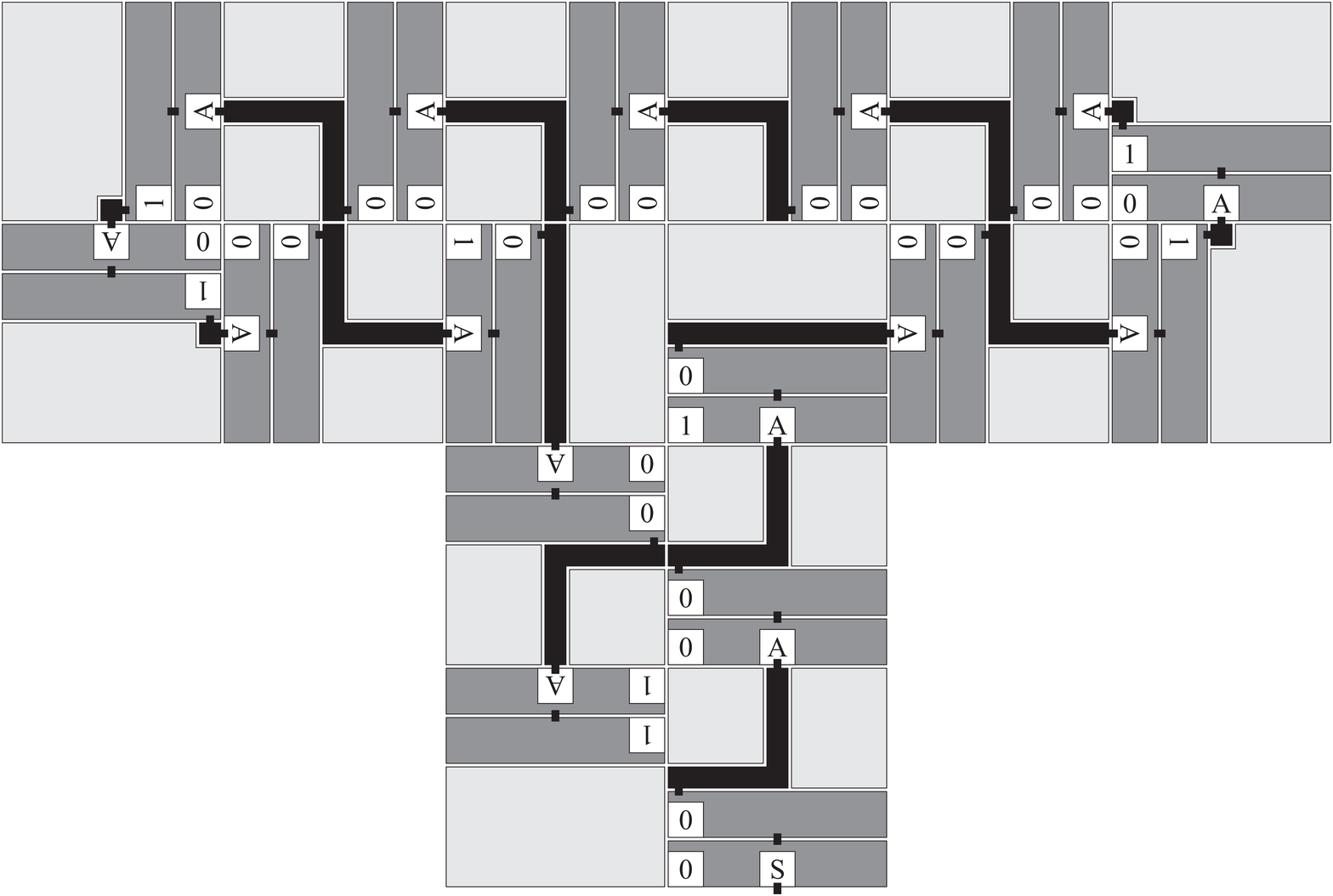}}
        \caption{\small \label{fig:secondconstruction} Overview of second construction for Theorem~\ref{thirdmaintheorem}.}
    \end{center}
    \vspace{-15pt}
\end{figure}

\begin{figure}[htp]%
\centering
    \subfloat[][Filler tiles. Here, $i \in \{ -6, \ldots, -1, 2, \ldots, 12 \}$, and $1 \leq j \leq 4$]{%
        \label{fig:square_gadget_filler}%
        \includegraphics[width=2.67in]{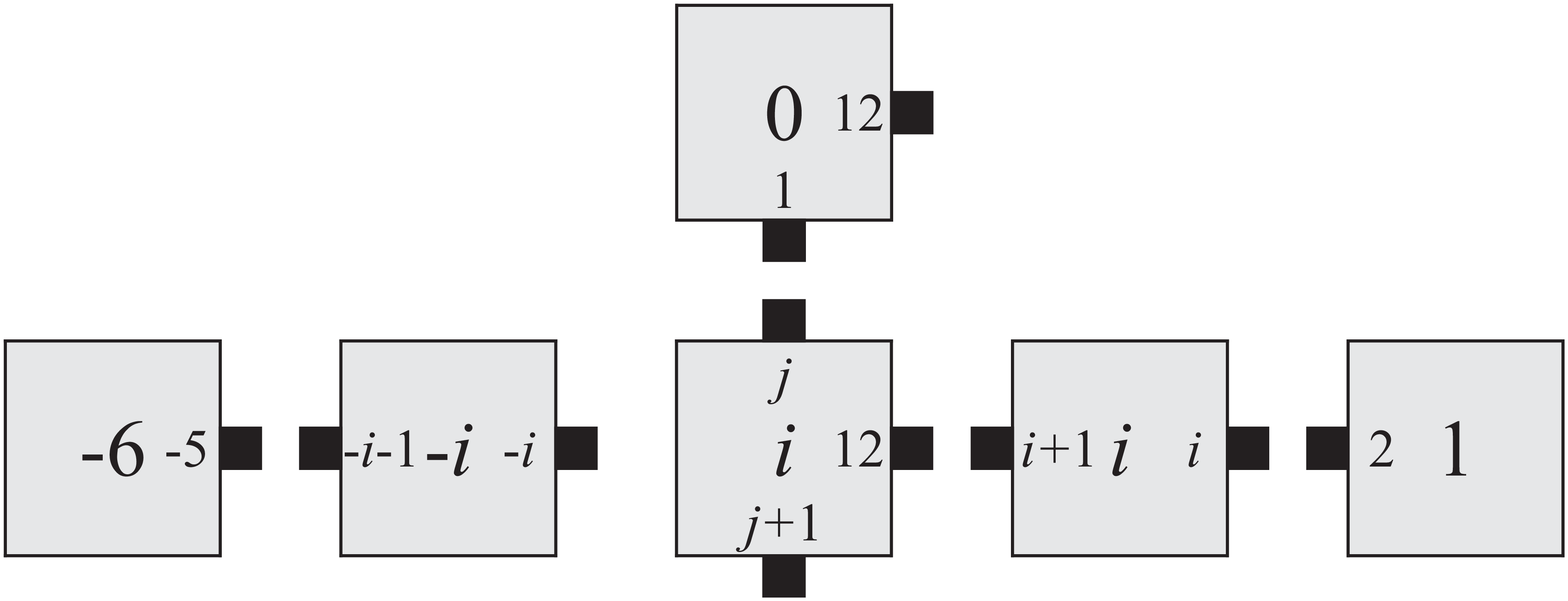}}
        \hspace{30pt}%
    \subfloat[][The `$\leftarrow$' tile initiates the growth of another square gadget ``west'' of the current square gadget; the other tile does not initiate any subsequent square gadgets]{%
        \label{fig:square_gadget_stop_and_left}%
        \includegraphics[width=2.16in]{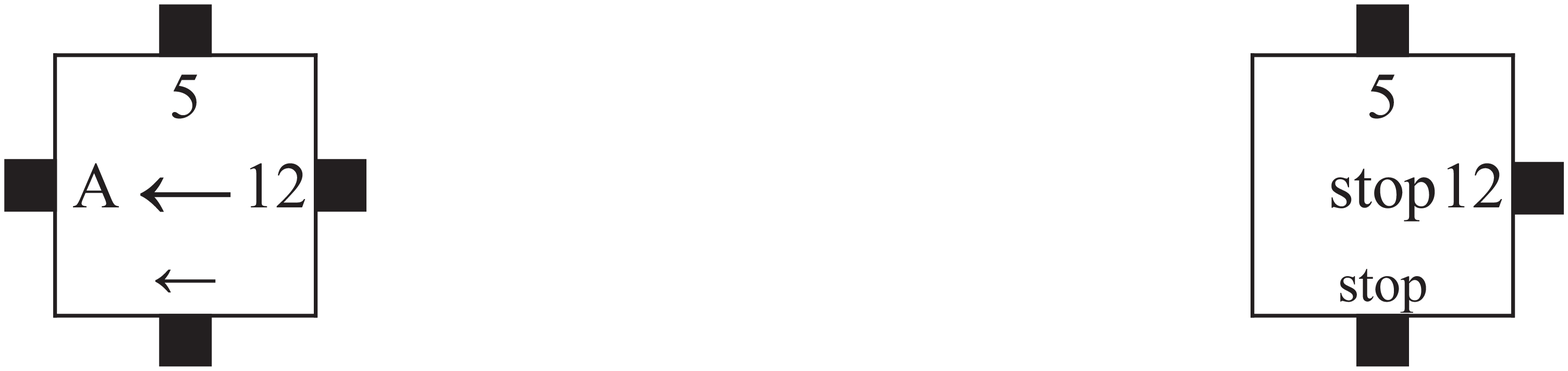}} \\
    \subfloat[][Tiles that initiate growth of another square gadget ``east'' of the current square gadget]{%
        \label{fig:square_gadget_right}%
        \includegraphics[width=6.07in]{square_gadget_right}}\\
    \subfloat[][Tiles that initiate growth of another square gadget directly ``north'' of the current square gadget]{%
        \label{fig:square_gadget_continue}%
        \includegraphics[width=2.74in]{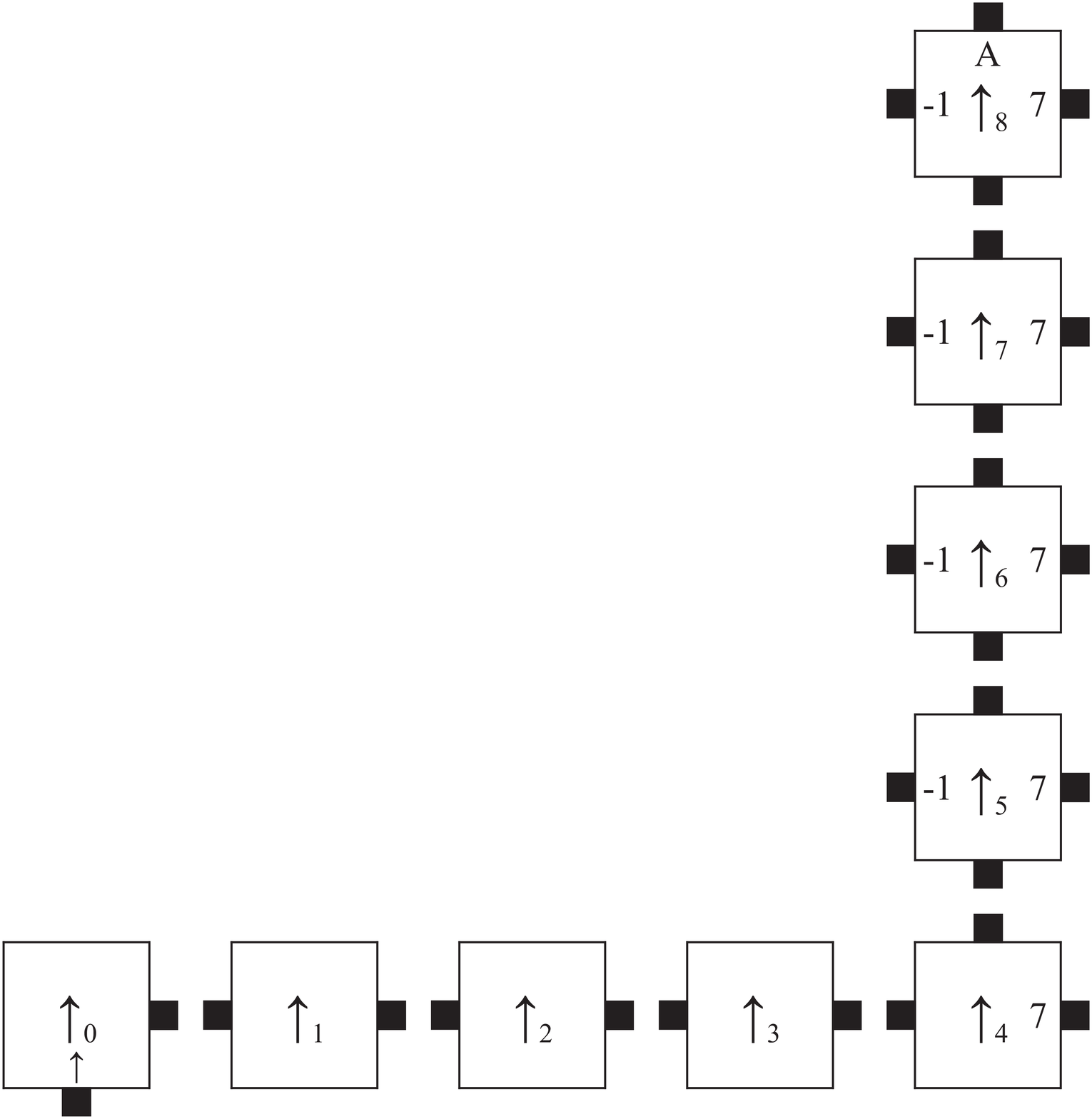}}
        \hspace{20pt}
    \subfloat[][Tiles for the second bit-flip gadget. Here $x,y \in \{0,1\}$, and $f: \{0,1\}^2 \longrightarrow \left\{ \uparrow,\leftarrow,\rightarrow,\textmd{stop}\right\}$ is defined as follows: $f(0,0) = \;\uparrow$, $f(0,1) = \;\leftarrow$, $f(1,0) = \;\rightarrow$, and $f(1,1) = \textmd{stop}$.]{%
        \label{fig:square_gadget_second_bitflip}%
        \includegraphics[width=3.25in]{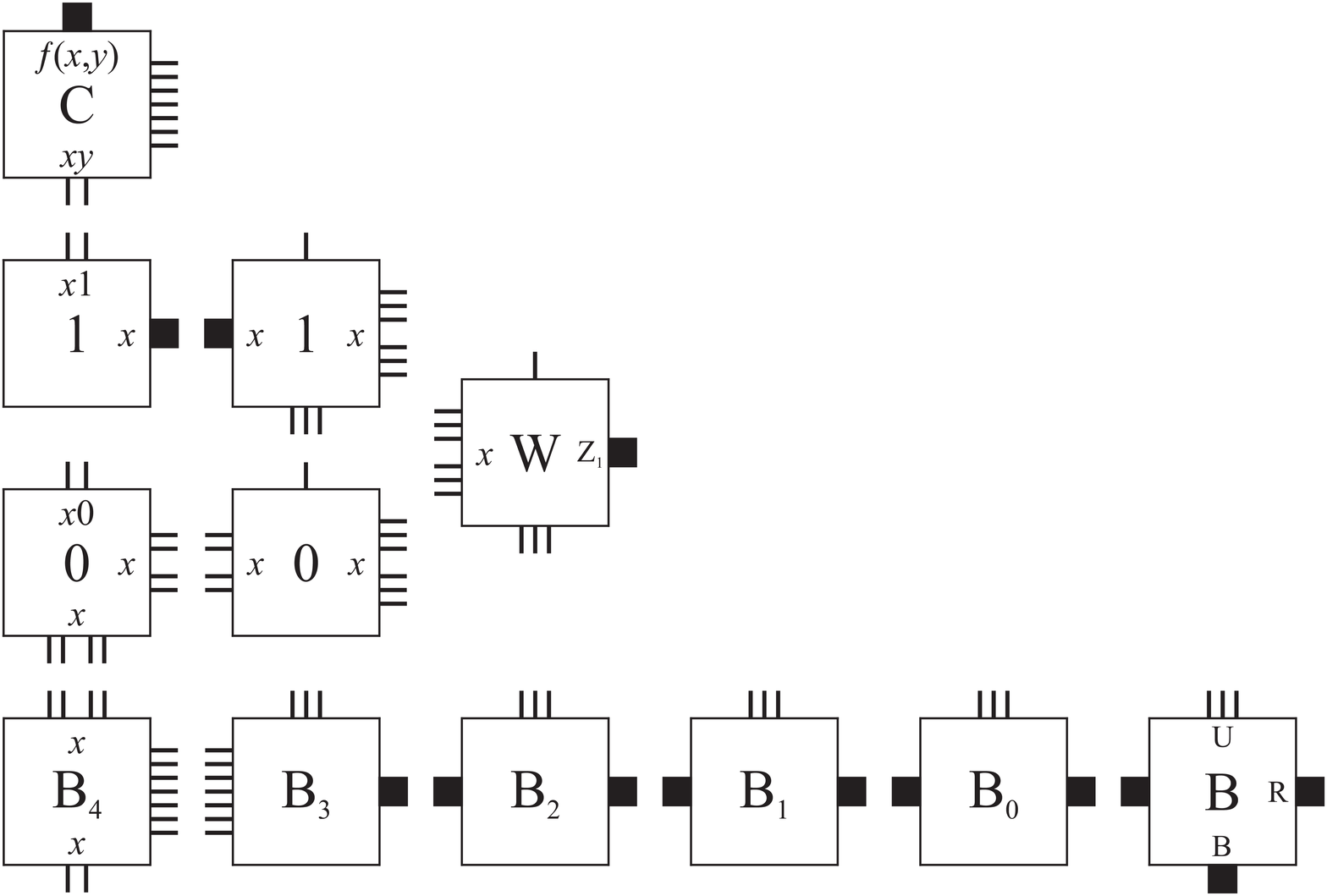}} \\
    \subfloat[][Tiles for the first bit-flip gadget. They are essentially the same as the those of the bit-flip gadget in \cite{KS07}]{%
        \label{fig:square_gadget_first_bitflip}%
        \includegraphics[width=6.0in]{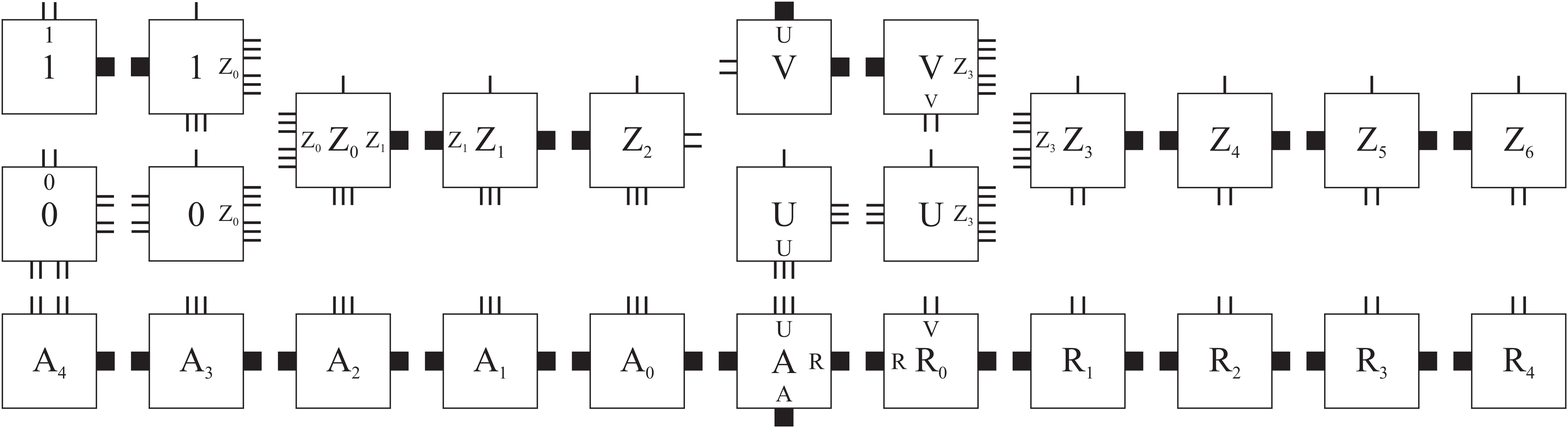}}
    \caption{\label{fig:tile_types_square_gadget} \small Tile types for the square gadget. The thick notches represent strength 9 bonds.}%
\end{figure}

\begin{proof}

Let $R = \left( \begin{array}{cc} 0 & -1 \\ 1 & 0 \end{array}\right)$ be the standard $90^\circ$ counterclockwise rotation matrix. If $t$ is a tile type, then we define $R(t) = t'$, such that, for all  $\vec{u} \in U_2$, $\textmd{ str}_{t'}\left(\vec{u}\right) = \textmd{str}_{t}\left(R\cdot \vec{u}\right)$ and $\textmd{col}_{t'}\left(\vec{u}\right) = \textmd{col}_{t}\left(R\cdot \vec{u}\right)$. Notice that $t'$ is simply the clockwise rotation of $t$. Let $t_{\textmd{seed}}$ be the single seed tile type defined in Figure~\ref{fig:seed_tile}.
\begin{figure}[htp]
    \begin{center}
        \includegraphics[width=0.50in]{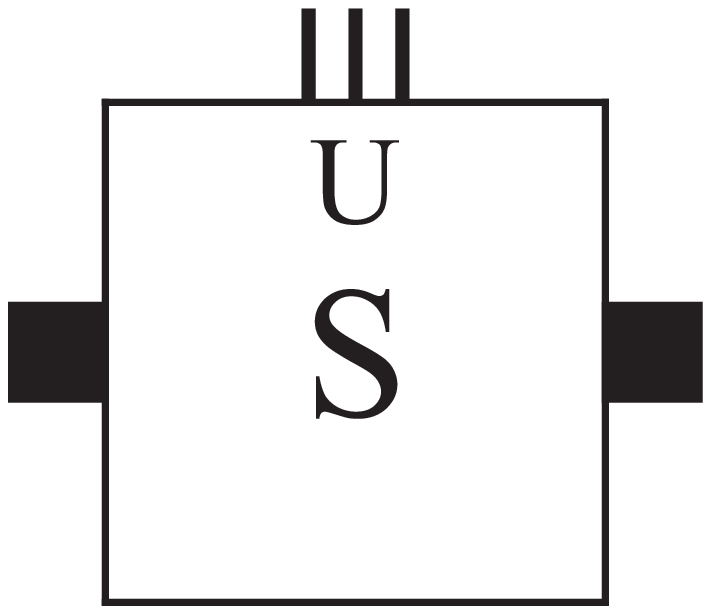}
        \caption{\label{fig:seed_tile} The unique seed tile type $t_{\textmd{seed}}$.}
    \end{center}
    \vspace{-10pt}
\end{figure}

Let $T$ be the set of tile types satisfying (1) $t_{\textmd{seed}} \in T$ and (2), for every tile type $t$ that is defined in Figure~\ref{fig:tile_types_square_gadget} in the technical appendix, $t, R(t), R^2(t)$ and $R^3(t)$ are all elements of $T$. This means that $T$ contains four logical copies of the square gadget each having a different ``type'' of direction. As a final technical matter, we adjust all of the glue colors (excluding the `A' glue color) on all of the tiles in $T$ such that two tiles can bind if and only if the respective square gadgets to which they belong have the same type of direction.

Let $X$ be an arbitrary finite shape such that there exists a Hamiltonian path $C = \left \langle \vec{v}_0, \vec{v}_1, \ldots, \vec{v}_{|X|-1} \right \rangle$ of $\fgg{X}$ (here, $C$ is a sequence of vertices). Let $\vec{u}_0 = (0,1)$. For all $1 \leq i < |X|$, define the unit vector $\vec{u}_i = v_{i} - v_{i-1}$. Define the temperature sequence  $\left \langle \tau_0, \tau_1, \ldots, \tau_{8\cdot |X|-1} \right \rangle$, where for each $0 \leq i < |X|-1$,
$\tau_{8i} = 4$,
$$
\tau_{8i+1} = \left\{
\begin{array}{ll}
4 & \textrm{ if } \vec{u}_i = R\cdot \vec{u}_{i-1} \textmd{ or } \vec{u}_i = \vec{u}_{i-1} \\
9 & \textrm{ otherwise},
\end{array} \right.
$$
$\tau_{8i+2} = 3$, $\tau_{8i+3} = 8$, $\tau_{8i+4} = 4$,
$$
\tau_{8i+5} = \left\{
\begin{array}{ll}
4 & \textrm{ if } \vec{u}_i = (-R)\cdot \vec{u}_{i-1} \textmd{ or } \vec{u}_i = \vec{u}_{i-1} \\
9 & \textrm{ otherwise,}
\end{array} \right.
$$
$\tau_{8i+6} = 3$, and $\tau_{8i+7} = 8$. Finally, let $\tau_{8|X|-8} = 4$, $\tau_{8|X|-7} = 9$, $\tau_{8|X|-6} = 3$, $\tau_{8|X|-5} = 8$, $\tau_{8|X|-4} = 4$, $\tau_{8|X|-3} = 9$, $\tau_{8|X|-2} = 3$, and $\tau_{8|X|-1} = 8$.
Let $\mathcal{T}^{X^{11}} = \left( T,\sigma,\left \langle \tau_i \right \rangle_{i=0}^{8\cdot |X|-1}\right)$. We will now describe how the scaled shape $X^{11}$ self-assembles in $\mathcal{T}^{X^{11}}$.

Assume without loss of generality that if $i = 0$, then $\tau_2 = \tau_6 = 4$ (i.e., $v_1 - v_0 = (0,1)$). The initial temperature subsequence $\left \langle \tau_0,\tau_1,\ldots \tau_7 \right \rangle$ assembles (1) an initial $11 \times 11$ ``seed'' square gadget, denoted $S_0$, from the single seed tile, and (2) the first $1 \times 11$ row (column) of an appropriately-rotated square gadget attached to the top side of the seed gadget.

In general, for each $0 < i < |X|-1$, the temperature subsequence $\left \langle \tau_{8i},\ldots \tau_{8i+7} \right \rangle$ assembles (1) an appropriately-rotated $11 \times 11$ square gadget, denoted $S_i$, attached to the square gadget assembled in the previous temperature subsequence, and (2) the first $1 \times 11$ row (column) of an appropriately-rotated square gadget attached to: the top side of $S_i$ if $\tau_{8i+1} = \tau_{8i+5} = 4$; the left side if $\tau_{8i+1} = 4$ and $\tau_{8i+5} = 9$; or the right side if $\tau_{8i+1} = 9$ and $\tau_{8i+5} = 4$. Intuitively, by our choices of $\tau_{8i+1}$ and $\tau_{8i+5}$, we can force the ``direction'' of self-assembly to follow the Hamiltonian path $C$. Namely, we set: $\tau_{8i+1} = \tau_{8i+5} = 4$ in order to \emph{continue} straight (relative to the current direction of self-assembly); $\tau_{8i+1} = 4$ and $\tau_{8i+5} = 9$ to initiate a relative \emph{left turn}; $\tau_{8i+1} = 9$ and $\tau_{8i+5} = 4$ to initiate a relative \emph{right turn}; and $\tau_{8i+1} = 9$ and $\tau_{8i+5} = 9$ to \emph{halt} self-assembly.

Finally, we have $\tau_{8|X| -7} = \tau_{8|X|-3} = 9$, and the last temperature subsequence assembles the final $11 \times 11$ (halting) square gadget attached to the square gadget assembled in the previous temperature subsequence.

Since each square gadget is an $11 \times 11$ square, and $C$ is a Hamiltonian path of $\fgg{X}$, we have that $X^{11}$ self-assembles in $\mathcal{T}^{X^{11}}$. An intuitive illustration of this process is depicted in Figures~\ref{fig:secondconstruction}\subref{fig:secondconstruction1} and~\ref{fig:secondconstruction}\subref{fig:secondconstruction2}. Also, Figures~\ref{fig:example} and~\ref{fig:square8} in the technical appendix give a detailed example of how to program a square gadget to turn left. A formal proof of the fact that $\mathcal{T}^{X^{11}}$ uniquely produces $X^{11}$ is, although tedious, straightforward, and therefore omitted.
\end{proof}

In our second construction, we will encode a Hamiltonian path of a particular finite shape $X$ into a temperature sequence in order to assemble the scaled-up version of $X$. Unfortunately, not all shapes have Hamiltonian paths, which might suggest that this approach is doomed to fail. Lucky for us, however, we have the following technical lemma.

\begin{lemma}
\label{ham_cycle_lemma}
If $X$ is a finite shape, then there exists a Hamiltonian cycle $C$ in $\fgg{X^2}$.
\end{lemma}
\begin{proof}
Note that, in this proof, we will think of Hamiltonian cycles as sequences of edges. For every finite shape $X$, define the set $\mathcal{B}(X) = \left\{ (x,y) \in X \mid \left(\exists \vec{u} \in U_2\right) \textmd{ satisfying } (x,y)+\vec{u} \not \in X \right\}$. One can think of the set $\mathcal{B}(X)$ as the set of all points from which it is possible to ``get away from'' the shape $X$ in one step. In what follows, we will prove that there exists a Hamiltonian cycle $C$ in $\fgg{X^2}$ with the following property $P$:
\begin{center}
    \parbox{5.0in}{For every $(w,x),(y,z) \in \mathcal{B}\left(X^2\right)$ with $(y,z) - (w,x) \in U_2$, such that $\left( \left \lfloor \frac{y}{2} \right \rfloor , \left \lfloor \frac{z}{2} \right \rfloor \right) = \left( \left \lfloor \frac{w}{2} \right \rfloor, \left \lfloor \frac{x}{2} \right \rfloor \right)$, the edge $((w,x),(y,z)) \in C$.}\\
\end{center}
Our proof is by induction on $|X|$. For the base case, we have $|X| = 1$, and it is routine to verify that $\fgg{X^2}$ has a Hamiltonian cycle $C$ satisfying property $P$.

For the inductive case, let $X$ be a shape with $|X| = k+1$. We will show that $\fgg{X^2}$ has a Hamiltonian cycle $C$ satisfying property $P$. Let $\vec{x} \in X$ be an arbitrary point such that $X - \{\vec{x}\}$ is a shape. Then define the shape $Y = X - \{\vec{x}\}$. Since $Y$ is a shape with $|Y| = k$, the induction hypothesis tells us that $\fgg{Y^2}$ has a Hamiltonian cycle $D = \left \langle e_0, e_1, \ldots e_{k-1} \right \rangle$ satisfying property $P$. We will use $D$ to construct a Hamiltonian cycle $C$ in $\fgg{X^2}$ having property $P$ as follows. Let $\vec{a},\vec{b},\vec{c},\vec{d} \in X^2$ be such that $\left\{ \vec{a}, \vec{b},\vec{c},\vec{d} \right\} = \left\{ (x,y) \; \left| \; \left( \left\lfloor \frac{x}{2} \right\rfloor, \left\lfloor \frac{y}{2} \right\rfloor \right) = \vec{x} \right.\right\}$. Since $D$ satisfies property $P$, there exist points $\vec{p},\vec{q} \in Y^2$ such that, there exists $\vec{u} \in U_2$ satisfying $\vec{a}+\vec{u} = \vec{p}$, $\vec{d}+\vec{u} = \vec{q}$, and $\left(\vec{p},\vec{q}\right) \in D$. Suppose that $e_j = \left( \vec{p},\vec{q}\right)$ for some $0 \leq j < k$. Finally, let $C = \left \langle e_0, e_1, \ldots, e_{j-1}, \left( \vec{p},\vec{a} \right), \left( \vec{a},\vec{b}\right), \left( \vec{b},\vec{c}\right), \left( \vec{c},\vec{d}\right), \left( \vec{d},\vec{q}\right), e_{j+1},\ldots,e_{k-1} \right \rangle$. It is clear that $C$ is a Hamiltonian cycle in $\fgg{X^2}$ having property $P$.
\end{proof}

Lemma~\ref{ham_cycle_lemma} says that, if we are comfortable with doubling the scaling factor, we can always use a Hamiltonian cycle to direct the self-assembly of an arbitrary scaled shape.

\begin{theorem}
\label{thirdmaintheorem}
There exists a tile set $T$ with $|T| = O(1)$ such that, for every finite shape $X$, there exists a temperature sequence $\left \langle \tau_i \right\rangle_{i=0}^{k-1}$ with $k = O(|X|)$, such that $\mathcal{T}^{X^{22}} = \left(T,\sigma,\left \langle \tau_i \right\rangle_{i=0}^{k-1}\right)$ uniquely produces $X^{22}$.
\end{theorem}

\begin{proof}
The theorem follows by Theorem~\ref{assemble_ham_cycle}, Lemma~\ref{ham_cycle_lemma}, and the simple observation that, for any finite shape $X$, $\left( X^2 \right)^{11} = X^{2\cdot 11} = X^{22}$.
\end{proof}

\section{Impossibility of Self-Assembly of Arbitrary Shapes with $O(1)$ Tile Types}

At this point, a natural question might be the following: Is the scaling factor in both of our constructions absolutely necessary? This question can be stated formally as follows.

\begin{question}[Kao and Schweller \cite{KS07}]
    \label{ks_open_question}
    Does there exist a tile set $T$, with $|T| = O(1)$, such that for every finite shape $X$, there exists a temperature sequence $\left \langle \tau_i \right \rangle_{i=0}^{k-1}$ such that $\mathcal{T} = \left( T,\sigma, \left \langle \tau_i \right \rangle_{i=0}^{k-1}\right)$ uniquely produces $X$?
\end{question}

In the remainder of this section, we prove that the answer to Question~\ref{ks_open_question} is ``no.'' In other words, we prove that the scaling factor in each of our constructions is necessary.

We must first, however, define some notation that we will use in our proof. If $\vec{\alpha} = \left(\alpha_0,\alpha_1,\ldots,\alpha_{l-1}\right)$ is an assembly sequence in $\mathcal{T}$ and $\vec{m} \in \mathbb{Z}^2$, then the $\vec{\alpha}$-{\it index} of $\vec{m}$ is $i_{\vec{\alpha}}(\vec{m}) = \min\{ i\in \mathbb{N} \; \left| \; \vec{m} \in \dom{\alpha_i} \right. \}$. That is, the $\vec{\alpha}$-index of $\vec{m}$ is the time at which any tile is first placed at location $\vec{m}$ by $\vec{\alpha}$. For each location $\vec{m} \in \bigcup_{0\leq i < l}{\dom{\alpha_{i}}}$, define $\textmd{IN}^{\vec{\alpha}}(\vec{m}) = \left\{ \vec{u} \in U_2 \; \left| \; i_{\vec{\alpha}}(\vec{m}+\vec{u}) < i_{\vec{\alpha}}(\vec{m}) \textmd{ and }
\textmd{str}_{\alpha_{i_{\vec{\alpha}}(\vec{m})}}(\vec{m},\vec{u})>0
\right.\right\}$. Intuitively, the set $\textmd{IN}^{\vec{\alpha}}(\vec{m})$ is the set of sides on which the \emph{first} tile that $\vec{\alpha}$ places at location $\vec{m}$ \emph{initially} binds. We now have the machinery to prove the following result.

\begin{theorem}
    \label{no_general_purpose_tile_system}
    For every tile set $T$, there exists a finite shape $X \subseteq \mathbb{Z}^2$ such that for each temperature sequence $\left \langle \tau_i \right \rangle_{i=0}^{k-1}$, $\mathcal{T} = \left( T,\sigma, \left \langle \tau_i \right \rangle_{i=0}^{k-1}\right)$ does not uniquely produce $X$.
\end{theorem}

\begin{proof}
Fix a set of tile types $T$, and let $X = \{0,\ldots, |T|\} \times \{0\}$. We will show that, given any temperature sequence $\left \langle \tau_i \right\rangle_{i=0}^{k-1}$, the tile system ${\mathcal T} = \left(T, \sigma, \left\langle \tau_i \right\rangle_{i=0}^{k-1}\right)$ does not uniquely produce $X$. To get a contradiction, assume that $\mathcal{T}$ uniquely produces $X$ and let $\alpha$ be the unique assembly satisfying $\alpha \in \termasm{\mathcal{T}}$. Since $|X| > |T|$, there must exist $\vec{x}_p,\vec{x}_q \in X$ such that $\alpha\left(\vec{x}_p\right) = \alpha\left(\vec{x}_q\right)$. Since $\mathcal{T}$ uniquely produces $X$, we know that, by definition, every assembly sequence in $\mathcal{T}$ is finite. Therefore, it suffices to exhibit an infinite assembly sequence in $\mathcal{T}$. The next fact is an easy consequence of the pigeonhole principle.

\begin{fact}
\label{fact_zero}
If $\vec{\alpha} = \left(\alpha_0,\alpha_1,\ldots,\alpha_{m-1}\right)$ is an assembly sequence in $\mathcal{T}$ such that, for all $0 \leq i < m$, $\dom{\alpha_i} \subseteq X$ and $\dom{\alpha_{m-1}} = X$, then there exists an infinite assembly sequence $\vec{\alpha}'$ in $\mathcal{T}$.
\end{fact}

For each assembly sequence $\vec{\alpha} = \left(\alpha_0,\alpha_1,\ldots,\alpha_{m-1}\right)$ in $\mathcal{T}$ such that for some $0 \leq i < m$, $\dom{\alpha_i} - X \ne \emptyset$, let $\vec{y}_{\vec{\alpha}}$ be the unique point $\vec{y}_{\vec{\alpha}} \not \in X$ such that, for all $\vec{z} \not \in X \cup \left\{\vec{y}_{\vec{\alpha}}\right\}$, $i_{\vec{\alpha}}\left(\vec{y}_{\vec{\alpha}}\right) < i_{\vec{\alpha}}\left(\vec{z}\right)$. Intuitively, the point $\vec{y}_{\vec{\alpha}}$ is the location of the first tile that $\vec{\alpha}$ places at any point not in $X$. The next fact gives sufficient conditions for the existence of an infinite assembly sequence in $\mathcal{T}$.

\begin{fact}
\label{fact_one}
Let $\vec{\alpha} = \left(\alpha_0,\alpha_1,\ldots,\alpha_{m-1}\right)$ be an assembly sequence in $\mathcal{T}$ such that for some $0 \leq i < m$, $\dom{\alpha_i} - X \ne \emptyset$, and $\vec{u}$ be the unique vector satisfying $\vec{u} \in \textmd{IN}^{\vec{\alpha}}(\vec{y}_{\vec{\alpha}})$. If $\textmd{str}_{\alpha_{i_{\vec{\alpha}}}\left(\vec{y}_{\vec{\alpha}}\right)}\left(\vec{y}_{\vec{\alpha}},\vec{u}\right) < \tau_{k-1}$ and, for every $\vec{x} \in X$,
$$
i_{\vec{\alpha}}\left(\vec{x}\right) < i_{\vec{\alpha}}\left(\vec{y}_{\vec{\alpha}}\right) \Rightarrow \alpha_{i_{\vec{\alpha}}\left(\vec{x}\right)}\left(\vec{x}\right) = \left\{ \begin{array}{ll} \alpha\left(\vec{x}-\left(\vec{x}_q - \vec{x}_p \right)\right) & \textmd{ if } \vec{x} - \vec{x}_q \in X \\ \alpha\left(\vec{x}\right) & \textmd{ otherwise,} \end{array}\right.
$$
then there exists an infinite assembly sequence $\vec{\alpha}'$ in $\mathcal{T}$.
\end{fact}
\begin{proof}(\emph{of Fact~\ref{fact_one}})
Let $(x_0,0) = \vec{x} \in X$ be the location of the rightmost tile in the assembly $\alpha_{i_{\vec{\alpha}}\left(\vec{y}_{\vec{\alpha}}\right)}$. We can define an \emph{infinite} assembly sequence $\vec{\alpha}' = \left( \alpha'_0, \alpha'_1, \ldots \right)$ as follows. For $0 \leq i \leq i_{\vec{\alpha}}\left(\vec{y}_{\vec{\alpha}}\right)$, let $\alpha'_i = \alpha_i$. Notice that at this point, the temperature of the current temperature phase must be less than $\tau_{k-1}$ since $\textmd{str}_{\alpha_{i_{\vec{\alpha}}}\left(\vec{y}_{\vec{\alpha}}\right)}\left(\vec{y}_{\vec{\alpha}},\vec{u}\right) < \tau_{k-1}$. This means that we can keep placing tiles sequentially at points in $X$ (because $\alpha$ is $\tau_{k-1}$-stable, whence every bond between adjacent tiles in $\alpha$ must be at least this strong). For $i_{\vec{\alpha}}\left(\vec{y}_{\vec{\alpha}}\right) \leq i < (|X|-1)+i_{\vec{\alpha}}\left(\vec{y}_{\vec{\alpha}}\right)-x_0$, let $\alpha'_{i+1}$ be the assembly obtained from $\alpha'_{i}$ by placing a tile of type $\alpha\left(\vec{x}+\left(i-i_{\vec{\alpha}}\left(\vec{y}_{\vec{\alpha}}\right)+1,0\right)-\left(\vec{x}_q - \vec{x}_p\right)\right)$ at the point $\vec{x}+\left(i-i_{\vec{\alpha}}\left(\vec{y}_{\vec{\alpha}}\right)+1,0\right)$ if $\vec{x}+\left(i-i_{\vec{\alpha}}\left(\vec{y}_{\vec{\alpha}}\right)+1,0\right) - \vec{x}_q \in X$, and $\alpha\left(\vec{x}+\left(i-i_{\vec{\alpha}}\left(\vec{y}_{\vec{\alpha}}\right)+1,0\right)\right)$ otherwise. At this point, a tile has been placed at every point in $X$. But why stop there? For $i \geq (|X|-1)+i_{\vec{\alpha}}\left(\vec{y}_{\vec{\alpha}}\right)-x_0-1$, let $\alpha'_{i+1}$ be obtained from $\alpha'_i$ by placing a tile of type $\alpha'_i\left(\left( i+1-\left(  i_{\vec{\alpha}}\left(\vec{y}_{\vec{\alpha}}\right)-x_0 -1 \right),0 \right)-\left( \vec{x}_q - \vec{x}_p \right)\right)$ at the point $\left( i+1-\left( i_{\vec{\alpha}}\left(\vec{y}_{\vec{\alpha}}\right)-x_0 -1 \right),0 \right)$. A routine induction argument can be used to show that $\vec{\alpha}'$ is an infinite (eventually) periodic assembly sequence in $\mathcal{T}$.
\end{proof}

\begin{fact}
\label{fact_two}
There exists an assembly sequence $\vec{\alpha}^* = \left( \alpha_0^*, \alpha_1^*, \ldots, \alpha_{m-1}^* \right)$ in $\mathcal{T}$ such that for some $0 \leq i < m$, $\dom{\alpha^*_i} - X \ne \emptyset$ and, for every $\vec{x} \in X$,
$$
i_{\vec{\alpha}^*}\left(\vec{x}\right) < i_{\vec{\alpha}^*}\left(\vec{y}_{\vec{\alpha}^*}\right) \Rightarrow \alpha_{i_{\vec{\alpha}^*}\left(\vec{x}\right)}\left(\vec{x}\right) = \left\{ \begin{array}{ll} \alpha\left(\vec{x}-\left(\vec{x}_q - \vec{x}_p \right)\right) & \textmd{ if } \vec{x} - \vec{x}_q \in X \\ \alpha\left(\vec{x}\right) & \textmd{ otherwise.} \end{array}\right.
$$
\end{fact}

Fact~\ref{fact_two} follows by the simple observations that every assembly sequence in $\mathcal{T}$ is finite, and until some tile is placed at some point outside of $X$, every tile must bind via a single input side.

Let $\vec{\alpha}^* = \left( \alpha_0^*, \alpha_1^*, \ldots, \alpha_{m-1}^* \right)$ be the assembly sequence in $\mathcal{T}$ given in Fact~\ref{fact_two}. Suppose that the first tile that $\vec{\alpha}^*$ places at $\vec{y}_{\vec{\alpha}^*} \not \in X$ binds to the single seed tile. In this case, the tile that is placed at $\vec{y}_{\vec{\alpha}^*}$ must bind via a single bond having strength less than $\tau_{k-1}$, because otherwise (if it were to bind with strength at least $\tau_{k-1}$), it would be possible to place a tile at $\vec{y}_{\vec{\alpha}^*}$ before finishing the final temperature phase (contradicting the the fact that $\mathcal{T}$ uniquely produces $X$). If the first tile that $\vec{\alpha}^*$ places at $\vec{y}_{\vec{\alpha}^*}$ binds to the single seed tile with strength less than $t_{k-1}$, then Fact~\ref{fact_one} gives us an infinite assembly sequence $\vec{\alpha}'$ in $\mathcal{T}$ - a contradiction.

Therefore, the first tile that $\vec{\alpha}^*$ places at the point $\vec{y}_{\vec{\alpha}^*}$ \emph{cannot} bind to the seed tile. Let $\vec{u}$ be the unique vector satisfying $\vec{u} \in \textmd{IN}^{\vec{\alpha}^*}(\vec{y}_{\vec{\alpha}^*})$. Suppose that $\textmd{str}_{\alpha_{i_{\vec{\alpha}^*}}\left(\vec{y}_{\vec{\alpha}^*}\right)}\left(\vec{y}_{\vec{\alpha}^*},\vec{u}\right) \geq \tau_{k-1}$. If $\left(\vec{y}_{\vec{\alpha}^*}+\vec{u}\right) - \vec{x}_q \not \in X$, then it would be possible to place some tile at the point $\vec{y}_{\vec{\alpha}^*} + \vec{u}$ during the final temperature phase because $\alpha_{i_{\vec{\alpha}^*}\left(\vec{y}_{\vec{\alpha}^*}+\vec{u}\right)}\left(\vec{y}_{\vec{\alpha}^*}+\vec{u}\right) = \alpha\left(\vec{y}_{\vec{\alpha}^*}+\vec{u}\right)$ (a contradiction). On the other hand, if $\left(\vec{y}_{\vec{\alpha}^*}+\vec{u}\right) - \vec{x}_q \in X$, then it would be possible to place some tile at the point $\left(\vec{y}_{\vec{\alpha}^*} + \vec{u}\right) - \left(\vec{x}_q - \vec{x}_p \right)$ during the final temperature phase because $\alpha_{i_{\vec{\alpha}^*}\left(\vec{y}_{\vec{\alpha}^*}+\vec{u}\right)}\left(\vec{y}_{\vec{\alpha}^*}+\vec{u}\right) = \alpha\left(\left(\vec{y}_{\vec{\alpha}^*} + \vec{u}\right)-\left(\vec{x}_q - \vec{x}_p \right)\right)$ (a contradiction).
Therefore, it must be that $\textmd{str}_{\alpha_{i_{\vec{\alpha}^*}}\left(\vec{y}_{\vec{\alpha}^*}\right)}\left(\vec{y}_{\vec{\alpha}^*},\vec{u}\right) < \tau_{k-1}$, and Fact~\ref{fact_one} gives us an infinite assembly sequence $\vec{\alpha}'$ in $\mathcal{T}$ - a contradiction.
\end{proof}

\section{Conclusion}
In this paper, we showed how to reduce the tile complexity for the self-assembly of arbitrary scaled shapes in the multiple temperature model. We developed two general purpose tile sets capable of assembling scaled-up versions of arbitrary finite shapes through appropriately chosen sequences of non-negative integer temperatures. While our first construction assembled shapes via short (asymptotically Kolmogorov-optimum) temperature sequences, the scaling factor grows (unboundedly) with the size of the shape being assembled. In contrast, our second construction assembled shapes via long temperature sequences but with a constant ``universal'' scaling factor in the sense that it is the same for every shape. We then proved that, for every constant-size tile set $T$, there is some finite shape that $T$ cannot uniquely produce via any temperature sequence, which implies the necessity of the scaling factor in both of our constructions. A natural direction for future theoretical research in the multiple temperature is the next question, which asks whether we can simultaneously optimize the two criteria of a short temperature sequence and a constant scaling factor, which were optimized separately in the constructions of this paper.
\begin{question}
Does there exist a tile set $T$ with $|T| = O(1)$ and $c \in \mathbb{N}$, such that, for every shape $X$, there exists a temperature sequence $\left \langle \tau_i \right\rangle_{i=0}^{k-1}$ with $k = O(K(X))$, such that $\mathcal{T}^{X^c} = \left(T,\sigma,\left \langle \tau_i \right\rangle_{i=0}^{k-1}\right)$ uniquely produces $X^c$?
\end{question} 

\subsubsection*{Acknowledgment}
This paper is a direct result of stimulating conversations with David Doty and Matthew Patitz.

\clearpage

\bibliographystyle{amsplain}
\bibliography{main,tam}

\providecommand{\bysame}{\leavevmode\hbox to3em{\hrulefill}\thinspace}
\providecommand{\MR}{\relax\ifhmode\unskip\space\fi MR }
\providecommand{\MRhref}[2]{%
  \href{http://www.ams.org/mathscinet-getitem?mr=#1}{#2}
}
\providecommand{\href}[2]{#2}
\begin{thebibliography}{10}

\bibitem{AdChGoHu01}
Leonard Adleman, Qi~Cheng, Ashish Goel, and Ming-Deh Huang, \emph{Running time
  and program size for self-assembled squares}, STOC '01: Proceedings of the
  thirty-third annual ACM Symposium on Theory of Computing (New York, NY, USA),
  ACM, 2001, pp.~740--748.

\bibitem{AGKS05}
Gagan Aggarwal, Qi~Cheng, Michael~H. Goldwasser, Ming-Yang Kao, Robert~T.
  Schweller, and Pablo~Moisset de~Espan\'{e}s, \emph{Complexities for
  generalized models of self-assembly}, SIAM Journal on Computing 34 (2005),
  1493--1515.

\bibitem{BarSchRotWin09}
Robert~D. Barish, Rebecca Schulman, Paul~W. Rothemund, and Erik Winfree,
  \emph{An information-bearing seed for nucleating algorithmic self-assembly},
  Proceedings of the National Academy of Sciences (2009).

\bibitem{BeckerRR06}
Florent Becker, Ivan Rapaport, and Eric R{\'e}mila, \emph{Self-assembling
  classes of shapes with a minimum number of tiles, and in optimal time},
  Foundations of Software Technology and Theoretical Computer Science (FSTTCS),
  2006, pp.~45--56.

\bibitem{ChenGoel04}
Ho-Lin Chen and Ashish Goel, \emph{Error free self-assembly with error prone
  tiles}, Proceedings of the 10th International Meeting on DNA Based Computers,
  2004.

\bibitem{DDFIRSS07}
Erik~D. Demaine, Martin~L. Demaine, S{\'a}ndor~P. Fekete, Mashhood Ishaque,
  Eynat Rafalin, Robert~T. Schweller, and Diane~L. Souvaine, \emph{Staged
  self-assembly: nanomanufacture of arbitrary shapes with ${O}(1)$ glues},
  Natural Computing \textbf{7} (2008), no.~3, 347--370.

\bibitem{RSAES}
David Doty, \emph{Randomized self-assembly for exact shapes}, Proceedings of
  the 50th Annual IEEE Symposium on Foundations of Computer Science, IEEE,
  2009, to appear.

\bibitem{KS07}
Ming-Yang Kao and Robert~T. Schweller, \emph{Reducing tile complexity for
  self-assembly through temperature programming}, Proceedings of the 17th
  Annual ACM-SIAM Symposium on Discrete Algorithms (SODA 2006), Miami, Florida,
  Jan. 2006, pp. 571-580, 2007.

\bibitem{KaoSchS08}
\bysame, \emph{Randomized self-assembly for approximate shapes}, International
  Colloqium on Automata, Languages, and Programming (ICALP) (Luca Aceto, Ivan
  Damgård, Leslie~Ann Goldberg, Magnús~M. Halldórsson, Anna Ingólfsdóttir, and
  Igor Walukiewicz, eds.), Lecture Notes in Computer Science, vol. 5125,
  Springer, 2008, pp.~370--384.

\bibitem{jSSADST}
James~I. Lathrop, Jack~H. Lutz, and Scott~M. Summers, \emph{Strict
  self-assembly of discrete {S}ierpinski triangles}, Theoretical Computer
  Science \textbf{410} (2009), 384--405.

\bibitem{LiVitanyiIntro}
Ming Li and Paul Vitanyi, \emph{An introduction to kolmogorov complexity and
  its applications (texts in computer science)}, Springer, February 1997.

\bibitem{Reif99localparallel}
John~H. Reif, \emph{Local parallel biomolecular computation}, University of
  Pennsylvania, American Mathematical Society, 1999, pp.~217--254.

\bibitem{RotOrigami06}
Paul W.~K. Rothemund, \emph{Folding {D}{N}{A} to create nanoscale shapes and
  patterns}, Nature \textbf{440}, no.~7082, 297--302.

\bibitem{Roth01}
Paul W.~K. Rothemund, \emph{Theory and experiments in algorithmic
  self-assembly}, Ph.D. thesis, University of Southern California, December
  2001.

\bibitem{RotWin00}
Paul W.~K. Rothemund and Erik Winfree, \emph{The program-size complexity of
  self-assembled squares (extended abstract)}, STOC '00: Proceedings of the
  thirty-second annual ACM Symposium on Theory of Computing (New York, NY,
  USA), ACM, 2000, pp.~459--468.

\bibitem{RoPaWi04}
Paul~W.K. Rothemund, Nick Papadakis, and Erik Winfree, \emph{Algorithmic
  self-assembly of dna sierpinski triangles}, PLoS Biology \textbf{2} (2004),
  no.~12.

\bibitem{Seem82}
Nadrian~C. Seeman, \emph{Nucleic-acid junctions and lattices}, Journal of
  Theoretical Biology \textbf{99} (1982), 237--247.

\bibitem{SolWin07}
David Soloveichik and Erik Winfree, \emph{Complexity of self-assembled shapes},
  SIAM Journal on Computing \textbf{36} (2007), no.~6, 1544--1569.

\bibitem{Wang61}
Hao Wang, \emph{Proving theorems by pattern recognition -- {II}}, The Bell
  System Technical Journal \textbf{XL} (1961), no.~1, 1--41.

\bibitem{Wang63}
\bysame, \emph{Dominoes and the {AEA} case of the decision problem},
  Proceedings of the Symposium on Mathematical Theory of Automata (New York,
  1962), Polytechnic Press of Polytechnic Inst. of Brooklyn, Brooklyn, N.Y.,
  1963, pp.~23--55.

\bibitem{Winf98}
Erik Winfree, \emph{Algorithmic self-assembly of {D}{N}{A}}, Ph.D. thesis,
  California Institute of Technology, June 1998.

\bibitem{WinBek03}
Erik Winfree and Renat Bekbolatov, \emph{Proofreading tile sets: Error
  correction for algorithmic self-assembly.}, DNA (Junghuei Chen and John~H.
  Reif, eds.), Lecture Notes in Computer Science, vol. 2943, Springer, 2003,
  pp.~126--144.

\end{thebibliography}

\section{Technical Appendix}

\end{document}